\documentclass[pra,twocolumn,floatfix,aps]{revtex4}
\usepackage{amsmath}
\usepackage{makeidx}
\usepackage{amssymb}
\usepackage{graphicx}
\usepackage{gensymb}

\usepackage[usenames]{color}
\usepackage[normalem]{ulem}
\usepackage{float}
\usepackage{soul}

\usepackage{hyperref}
\usepackage[T1]{fontenc} 

\begin{document}
%%%%%%%%%%%%%%%%%%%%%%%%%%%%%%%%%%%%%%%%%%%%%%%%%%%%%%%%%%%%%%%%%%%%%%%%%%%%%%
%%%%                     Title and authors                                %%%%
%%%%%%%%%%%%%%%%%%%%%%%%%%%%%%%%%%%%%%%%%%%%%%%%%%%%%%%%%%%%%%%%%%%%%%%%%%%%%%

\title{Supersolid-like solitons in spin-orbit coupled spin-$2$ 
condensate}
\author{Pardeep Kaur\footnote{2018phz0004@iitrpr.ac.in}}
\author{Sandeep Gautam\footnote{sandeep@iitrpr.ac.in}}
\affiliation{Department of Physics, Indian Institute of Technology Ropar, Rupnagar, Punjab 140001, India}
\author{S. K. Adhikari\footnote{sk.adhikari@unesp.br, 
         http://www.ift.unesp.br/users/adhikari}}
\affiliation{Instituto de F\'{\i}sica Te\'orica, Universidade Estadual
             Paulista - UNESP,  01.140-070 S\~ao Paulo, S\~ao Paulo, Brazil}
      
%%%%%%%%%%%%%%%%%%%%%%%%%%%%%%%%%%%%%%%%%%%%%%%%%%%%%%%%%%%%%%%%%%%%%%%%%%%%%%
%%%%%%%%%%                    Abstract                             %%%%%%%%%%%
%%%%%%%%%%%%%%%%%%%%%%%%%%%%%%%%%%%%%%%%%%%%%%%%%%%%%%%%%%%%%%%%%%%%%%%%%%%%%%

%\date{\today}
\begin{abstract}
We study supersolid-like crystalline structures emerging in the stationary states 
of a quasi-two-dimensional spin-orbit (SO)-coupled spin-2 condensate  
in the ferromagnetic, cyclic, and antiferromagnetic phases by solving a
mean-field model. Interplay of different strengths of SO coupling and 
interatomic interactions gives rise to a variety of non-trivial density 
patterns in the emergent solutions. For small SO-coupling 
strengths $\gamma$ ($\gamma \approx 0.5$), the ground state is an axisymmetric  
multi-ring soliton for polar, cyclic and weakly-ferromagnetic 
interactions, whereas for stronger-ferromagnetic interactions a circularly-asymmetric 
soliton emerges as the ground state. Depending on the values of interaction 
parameters, with an increase in SO-coupling strength, a stripe phase may also emerge 
as the ground state for polar and cyclic interactions. For intermediate values of SO-coupling strength 
($\gamma \approx 1$), in addition to these solitons, one could have a  
quasi-degenerate triangular-lattice soliton in all magnetic phases. 
On further increasing the SO-coupling strength ($\gamma \gtrapprox 4$), a square-lattice and 
a superstripe soliton emerge as quasi-degenerate states. The 
emergence of all these solitons  can be inferred from a study of  solutions 
of the single-particle Hamiltonian. 
\end{abstract}

\maketitle

%%%%%%%%%%%%%%%%%%%%%%%%%%%%%%%%%%%%%%%%%%%%%%%%%%%%%%%%%%%%%%%%%%%%%%%%%%%%%%%%
%%%%%%%%%%%%%%%.                  Introduction
%%%%%%%%%%%%%%%%%%%%%%%%%%%%%%%%%%%%%%%%%%%%%%%%%%%%%%%%%%%%%%%%%%%%%%%%%%%%%%%%
\section{Introduction}
The experimental realization of spinor condensates has opened up a plethora of 
possibilities to explore the physics of these quantum degenerate gases 
\cite{spinor_review}. As the constituent atoms of the condensates are neutral, 
an exciting challenge in the field of quantum degenerate gases was to introduce 
the spin-orbit (SO) coupling, i.e., the coupling between the spin of an atom and
its linear momentum. The same was realized in experiments by creating 
non-abelian gauge potentials through Raman lasers which coherently couple 
the spin-component states of a spinor Bose-Einstein condensate (BEC) \cite{lin2011spin,campbell2016magnetic,Zhang,Wu}.
An SO-coupled BEC has been used as a quantum simulator to study spin-Hall
effect \cite{Beeler}, fractional topological insulators \cite{Levin}, and 
spin-current generation \cite{Li_spin_current}, etc. \cite{galitski2013spin}. 
In a pseudospin-1/2 ($F=1/2$) spinor condensate with an equal-strength mixture
of Rashba \cite{rashba-soc} and Dresselhaus \cite{dresselhaus-soc} SO
couplings,   three ground-state phases emerge, e.g. stripe, plane-wave, and zero-momentum 
phases \cite{Li-spin-1/2}. An equal-strength mixture of   Rashba and Dresselhaus SO 
couplings  has also been realised in spin-1 condensates 
\cite{campbell2016magnetic} and experimental schemes to realize Rashba 
SO-coupled spin-1 ($F=1$) and spin-2 ($F=2$) BECs have also been proposed \cite{Anderson}. 

A vector-bright  soliton is 
a self-bound multi-component solitary wave which 
maintains its shape while moving with a constant velocity.
In the absence of SO coupling, in quasi-two-dimensional (quasi-2D) \cite{townes}
and three-dimensional (3D) \cite{r1} settings a soliton cannot be stabilized due to a collapse instability.
However, 
it has been theoretically demonstrated \cite{Kartashov} that an SO-coupling leads to a
stabilization of self-trapped solutions like bright solitons in   quasi-2D \cite{gautam2017vortex}
and 3D \cite{gautam2018three} spinor BECs. Vector-bright 
solitons have been studied extensively in 
SO-coupled quasi-one-dimensional (quasi-1D) \cite{spin-1/2-Q1D}, quasi-2D 
\cite{spin-1/2-Q2D}, and 3D pseudospin-1/2 BECs 
\cite{spin-1/2-3D}. These self-trapped solitary waves have also been predicted 
to emerge in SO-coupled quasi-1D \cite{gautam2015mobile}, 
quasi-2D \cite{gautam2017vortex,adhikari2021multiring}, and 
3D spin-1 BECs \cite{gautam2018three}. In an SO-coupled 
quasi-2D spin-1 BEC, the existence of square-lattice solitons with  a square-lattice 
modulation in the total density has also been demonstrated   \cite{adhikari2021multiring}. 
However, these self-trapped solitons are still unexplored in the case of 
quasi-2D SO-coupled spin-2 condensates and we undertake a comprehensive study of the same in this paper.
A spin-2 BEC has three magnetic phases compared to two for a spin-1 BEC. 
The SO-coupled spin-2 BEC density is expected to exhibit more complex symmetry properties compared to the 
spin-1 case and  %To create an SO-coupled 
%BEC in a laboratory, laser beams coupling different spin component states 
%are needed. As the number of such spin components increases with the 
%spin of the BEC atoms, more laser beams are needed to create an SO-coupled
%spin-2 BEC, as compared to an SO-coupled spin-1 BEC, which will lead to a
%more complex electromagnetic trapping potential \cite{}.  
%In addition, a spin-2 BEC has three {\color{red}magnetic} phases compared to two for a spin-1 BEC. 
%Hence it seems reasonable that a  spin-2 SO-coupled BEC may host different 
%types of supersolid states not possible in the case of an SO-coupled spin-1 
%BEC.
the interplay of spin-independent and two spin-dependent interactions with SO coupling
is expected to lead to a richer variety of emergent patterns in a spin-2 BEC \cite{emergent_lattice, symmetries_lattice}. 
%Similarly, a spin-3 SO-coupled BEC, with seven spin components and 
%nine \cite{spin3} possible magnetic phases, will require  many more lasers to create 
%an SO-coupling.  Consequently, the solitons in an SO-coupled spin-3 BEC 
%will exhibit a diverse class of symmetry properties not possible 
%in a spin-1 or spin-2 SO-coupled BEC. 

 An exciting recent development in 
the field has been the experimental realization of a { supersolid} phase of 
matter in dipolar BECs, where by tuning the ratio of dipolar to contact 
interactions, the system first undergoes a phase transition to a
supersolid 
phase, which is followed by a crossover to an insulating phase with 
a further decrease in the strength of contact interaction \cite{Bottcher}. 
The excitation spectrum of the dipolar BECs in the supersolid phase has further 
confirmed that this phase corresponds to a  simultaneous (and spontaneous) 
breaking of continuous translational and global gauge symmetries 
\cite{Natale}. The existence of a supersolid-like stripe phase with 
both diagonal and off-diagonal orders has been observed in SO-coupled 
pseudospin-1/2 spinor condensates \cite{Ketterle}. As no additional symmetry 
is broken vis-\`a-vis the system without Raman coupling, these supersolid-like 
stripes have been termed  {superstripes} \cite{Li,Putra}. 

In the present study, we consider a quasi-2D spin-2 
condensate with a Rashba SO coupling. We investigate theoretically  the self-trapped
solitons of the BEC in the mean-field approximation 
\cite{spin2-phases-ciobanu,spin2-phases-ueda}, wherein a spin-2 condensate 
is described by a set of five coupled Gross-Pitaevskii (GP) equations. We study the 
soliton formation  for  small SO-coupling strength $\gamma$ 
($\gamma \approx 0.5$), for moderate SO-coupling strength ($\gamma \approx 1$) 
and also for large SO-coupling strengths ($\gamma \gtrapprox 4$). For small 
SO-coupling strengths, $(-2,-1,0,+1,+2)$-type
multi-ring solitons appear in the three possible magnetic phases $-$ ferromagnetic, 
anti-ferromagnetic, and  cyclic $-$  as the stationary-state solution, where the numbers 
in the parentheses are  the phase winding numbers in components $m_f=+2,+1,0,-1,-2$, respectively,
in addition to circularly-asymmetric solutions. We minimize the 
spin-dependent-interaction energy  and  the SO-coupling energy to
establish the allowed values of angular momentum (phase winding numbers) in different components of the system.

For moderate ($\gamma \approx 1$) to strong SO-coupling strengths ($\gamma \gtrapprox 4$),
depending on interaction parameters, multiple quasi-degenerate solitons 
may emerge  in different magnetic phases  which include triangular-lattice
soliton, with hexagonal lattice formation and { square-lattice soliton}, with square lattice
formation,  in addition to stripe and superstripe solitons,  all with supersolid-like properties.
These states are quasi-degenerate because of internal symmetry properties. 
%{\color{red}In the limit of vanishing attractive spin-independent interaction ($c_0$), all localized 
%solitonic states  will become  uniform homogeneous with the same energy. The different localized quasi-degenerate %solitonic states
%are created for a non-zero  $|c_0|$ and as $|c_0|$ increases the degeneracy will be gradually removed.}
In the limit of vanishing attractive interactions, all localized solitonic states will have the 
same energy. The different localized quasi-degenerate solitonic states are created for a non-zero $|c_0|$ and 
as $|c_0|$ increases the degeneracy will be gradually removed.
Of these different states, the triangular-lattice, square-lattice, and superstripe solitons have spatially-periodic
modulation in both component and total densities, whereas the stripe soliton does not have any 
modulation in total density. We also construct the degenerate ground state solutions of the 
non-interacting SO-coupled condensate in order to anticipate the 
different types of solitons with supersolid-like properties,  which might emerge with the introduction of 
interactions.
We confirm the stability of these solutions 
by real-time simulation  over long periods of time using the converged 
imaginary-time wave function as the initial state.

 The head-on collision between two $(-2,-1,0,+1,+2)$-type multi-ring solitons has also been studied. 
The {mean-field GP equations} in presence of an SO coupling {are} not Galilean invariant and we introduce a
Galilean-transformed GP {equations} to study a moving soliton.  We find that  a moving multi-ring soliton 
gets deformed with the increase of velocity and ceases to exist beyond a critical velocity; so we study the collision 
at small velocities. At larger velocities the collision is found to be quasi 
elastic  with the solitons passing through each other. At small velocities the collision is inelastic 
and the two solitons join to form a single entity and the identity of the colliding solitons is lost.

The paper is organized as follows. In the Sec. \ref{section2}, we 
present  the mean-field GP equations  of an SO-coupled spin-2 BEC. In Sec. \ref{section2a}, assuming a circular
symmetry we establish the allowed phase-winding numbers for the system
from a minimization of the interaction and SO-coupling energy terms. 
In Sec. \ref{section2b},  we  demonstrate the possibility of multi-ring, 
stripe, square-lattice, and triangular-lattice formation from solutions of 
the non-interacting system. In Sec. \ref{section3}, we discuss a variety of self-trapped solutions
emerging at different SO-coupling strengths from a numerical solution of the GP equations. 
In Sec. \ref{section3a}, we  show that axisymmetric multi-ring and circularly-asymmetric solitons,
the latter for ferromagnetic BECs, are possible for small SO-coupling strengths $\gamma$. In Sec. \ref{section3b}, we demonstrate that  
quasi-degenerate axisymmetric multi-ring and triangular-lattice solitons emerge for medium values of 
$\gamma$. In Sec. \ref{section3c}, we establish the formation of multi-ring, square-lattice,
superstripe and stripe solitons for large $\gamma$. We confirm the dynamic 
stability of the different solitons using real-time evolution with the addition of a small random noise 
to the order parameter in Sec. \ref{dyn_inst}. The bifurcation behaviour is discussed in Sec. \ref{bifur}. The Galilean transformed 
mean-field model for the condensate is introduced in Sec.
\ref{section3e}, which we use to study the moving  solitons and
collisions between them.
%In section (VI), we give a summary of our findings.

%%%%%%%%%%%%%%%%%%%%%%%%%%%%%%%%%%%%%%%%%%%%%%%%%%%%%%%%%%%%%%%%%%%%%%%%%%%%%%%%
%%%%%%%%%%%%                     Mean Field Model                        
%%%%%%%%%%%%%%%%%%%%%%%%%%%%%%%%%%%%%%%%%%%%%%%%%%%%%%%%%%%%%%%%%%%%%%%%%%%%%%%%

\section{Mean Field Model for Spin-Orbit-Coupled spin-2 BEC} 
\subsection{Gross-Pitaevskii equations}
\label{section2}

We consider an SO-coupled spin-2 spinor BEC free {in} the $x$-$y$ plane 
and confined by a harmonic trap ${V({\bf r})} = m{\omega_z}^2 z^2/2$ along the 
$z$-direction to its Gaussian ground state. The trapping frequency $\omega_z$ is strong enough to freeze the 
dynamics along $z$ direction. The single-particle Hamiltonian of this system in 
the presence of Rashba SO coupling is given by \cite{Zhai_reviews}
\begin{align}
    H_0 = \frac{p_x^2+p_y^2}{2m}  + \gamma (p_y S_x - p_x S_y),\label{sph}
\end{align}
where $p_x = -i\hbar \partial_x \equiv-i\hbar \partial/\partial x$ and $p_y = -i\hbar \partial_y\equiv -i\hbar\partial/\partial y$
are the momentum operators along $x$ and $y$ axes, respectively, $\gamma$ is the 
strength of SO coupling, $S_x$ and $S_y$ are the irreducible 
representations of the $x$ and $y$ components of angular momentum operators for 
spin-2 particle, respectively. The $(j',j)$th element of these $5\times 5$
matrices are
\begin{eqnarray}
(S_x)_{j',j} &=& \textstyle \frac{1}{2}\left(\textstyle \sqrt{(2-j)(2+j+1)}\hbar\delta_{j',j+1}\right.
                 \nonumber\\ 
             & &+\left.\sqrt{(2+j)(2-j+1)}\hbar\delta_{j',j-1}\right),\\
(S_y)_{j',j} &=& - \textstyle \frac{1}{2}i\left(\sqrt{(2-j)(2+j+1)}\hbar\delta_{j',j+1}\right.
                 \nonumber\\ 
             & &-\left.\sqrt{(2+j)(2-j+1)}\hbar\delta_{j',j-1}\right),
%(S_z)_{m',m} = m\hbar \delta_{m',m},
%\Sigma_x=\begin{pmatrix}
%0 & 1 & 0 & 0 & 0 \\
%1 & 0  & \sqrt{\frac{3}{2}} & 0 & 0\\
%0 & \sqrt{\frac{3}{2}}  & 0 & \sqrt{\frac{3}{2}} & 0\\
%0 & 0  & \sqrt{\frac{3}{2}} & 0 & 1\\
%0 & 0 & 0 & 1 & 0 \\
%\end{pmatrix}, \\
%\Sigma_y= i \begin{pmatrix}
%0 & -1 & 0 & 0 & 0 \\
%1 & 0  & -\sqrt{\frac{3}{2}} & 0 & 0\\
%0 & \sqrt{\frac{3}{2}}  & 0 & -\sqrt{\frac{3}{2}} & 0\\
%0 & 0  & \sqrt{\frac{3}{2}} & 0 & -1\\
%0 & 0 & 0 & 1 & 0 \\
%\end{pmatrix},
\end{eqnarray}
where $j'$ and $j$ represent the spin projections $m_f$ 
and can have values $\pm 2,\pm 1,0$.

The reduced quasi-2D spinor 
BEC can be described by a set of five coupled mean-field partial differential GP
equations for the wave-function components $\phi_j$ {and} are given in dimensionless form as \cite{spinor_review}
\begin{subequations}
\begin{eqnarray}
i \partial_t  \phi_{\pm 2} &=& \mathcal{H} \phi_{\pm 2} + 
c_0 {\rho} \phi_{\pm 2}+ c_1 (F_{\mp} \phi_{\pm 1} 
\pm 2 F_{z} \phi_{\pm 2})\nonumber\\  
&&+c_2 \textstyle  \frac{1}{\sqrt 5}{\Theta \phi_{\mp 2}^*}+ \Gamma_{\pm 2}, 
\label{cgpet3d-1}\\
i\partial_t \phi_{\pm 1} &=& \mathcal{H} \phi_{\pm 1} 
+ c_0 {\rho}\phi_{\pm 1} + c_1 \Big(\textstyle\sqrt{\frac{3}{2}} F_{\mp}
\phi_{0} +F_{\pm} \phi_{\pm 2}\nonumber\\ 
&&\pm F_{z} \phi_{\pm 1}\Big)
- c_2\textstyle \frac{1}{\sqrt 5}  {\Theta\phi_{\mp 1}^*} + \Gamma_{\pm 1}, 
\label{cgpet3d-2}\\
i\partial_t \phi_0 &=& \mathcal{H} \phi_0 
+ c_0 {\rho}
\phi_0 + c_1 {\textstyle\sqrt{\frac{3}{2}}}  (F_{-}\phi_{-1}+  
F_{+}\phi_{+1})\nonumber\\  
&&+ c_2 \frac{1}{\sqrt{5}} {\Theta \phi_{0}^*}+\Gamma_0, \label{cgpet3d-3}
\end{eqnarray}
\end{subequations}
where
\begin{align*}
\mathcal{H}&=-\textstyle\frac{1}{2} \left({\partial_x}^2 + 
              {\partial_y}^2\right), \\
%\quad 
%              \rho =\textstyle \sum_{j=-2}^2 |\phi_j|^2,\\
\Theta     &=\textstyle \frac{1}{\sqrt 5}  ({2\phi_{+2} \phi_{-2} - 2\phi_{+1}\phi_{-1}+ \phi_0^2}),
              \quad F_z =\textstyle \sum_j  j|\phi_j|^2, \\
F_- &= F_+^* = 2\phi_{-2}^* \phi_{-1} + \sqrt{6}\phi_{-1}^*\phi_0 + 
                \sqrt{6} \phi_0^* \phi_{+1} + 2 \phi_{+2} \phi_{+1}^* \, ,
\end{align*}
where $\partial _t \equiv \partial/\partial t,$  $\rho_j(x,y)= |\phi_j(x,y)|^2$ are component 
densities and $\rho(x,y)\equiv \sum_j \rho_j(x,y) $ is the total density,
$F_{\pm}=F_x\pm i F_y$,  $|{\bf F}|^2=F_x^2+F_y^2+F_z^2\equiv F_+F_-+F_z^2$,
where $F_x$, $F_y$, $F_z$ are the three components of the {spin-density} vector {\bf F},
and $\Theta$ is the spin-singlet pair amplitude.
In Eqs. (\ref{cgpet3d-1})-(\ref{cgpet3d-3}), the interaction parameters and 
SO coupling terms are defined as
\begin{subequations}
\begin{align}
\label{gammas}
c_0 =& \frac{2\sqrt{2\pi} N(4 a_2 +3 a_4)} {7 a_{\rm osc}},
       \: c_1= \frac{2\sqrt{2\pi} N(a_4-a_2)}{7 a_{\rm osc}},\\
c_2 =& \frac{2\sqrt{2\pi} N(7a_0-10a_2+3a_4 )}{7 a_{\rm osc}}\\
\Gamma_{\pm 2}=&-i\gamma\left({\partial_y}
\pm i\partial_x\right)\phi_{\pm 1},\\
\Gamma_{\pm 1}=&-i \textstyle \sqrt{\frac{3}{2}}\gamma\left({\partial_y}
                   \pm i{\partial_x}\right)\phi_0 
                - i\gamma\left({\partial_y} 
                   \mp i{\partial_x } \right) \phi_{\pm 2} , \\
\Gamma_{0}=&-i\textstyle\sqrt{\frac{3}{2}}\gamma[{(\partial_y   -i\partial_x )  \phi_{+1}} 
            + {(\partial_y+i\partial_x)\phi_{-1}} 
           ],
\end{align}
\end{subequations}
where $a_0$, $a_2$, and $a_4$ are $s$-wave scattering lengths in the possible total spin channels 
0, 2 and 4, respectively,  for a spin-2 BEC, and $N$ is the total number 
of bosons.

\begin{figure}[t]
\begin{center}
\includegraphics[trim = 0cm 0cm 0cm 0cm, clip,width=.7\linewidth,clip]{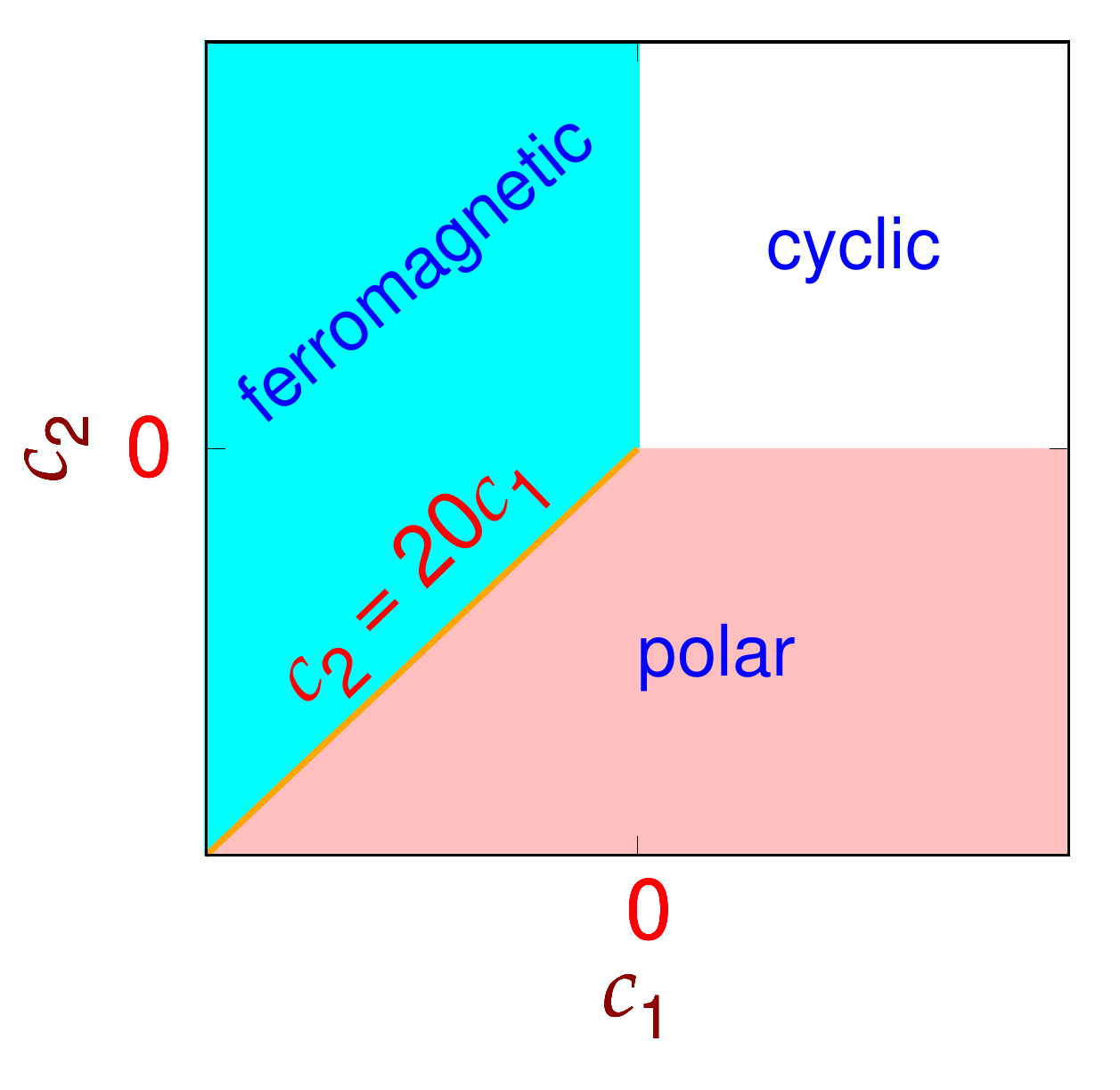}
\caption{(Color online) The $c_2$ versus $c_1$ phase plot (not to scale) illustrating ferromagnetic,
anti-ferromagnetic, and cyclic phases {in the absence of SO coupling}. All the figures
in this work are plotted in dimensionless units.}
\label{phase}
\end{center}
\end{figure}

In this study we will consider a self-attractive $(c_0<0)$  system.
Depending on the values of $c_1$ and $c_2$ we  can have three magnetic phases
\cite{spinor_review} $-$ ferromagnetic, anti-ferromagnetic, and cyclic $-$ as
illustrated in Fig. \ref{phase}.
The units of length, density, time, and energy considered in Eqs. 
(\ref{cgpet3d-1})-(\ref{cgpet3d-3}) are oscillator length $a_{\rm osc} =  
\sqrt{\hbar/m\omega_z}$, $a_{\rm osc}^{-2}$, $\omega_z^{-1}$, and 
$\hbar \omega_z$, respectively. The dimensionless formulation of mean-field 
model for the condensate has  the normalization 
condition   $\int\rho(x,y) dxdy =  1$. The number of particles  along with energy
\begin{eqnarray}
E &=& \textstyle \int dxdy
\left[\textstyle \sum_{j=-2}^{+2}\phi_j^{*}{\cal H}\phi_j
+ \frac{1}{2}c_0{\rho}^2+ \frac{1}{2}c_1|{\bf F}|^2 \right.\nonumber\\ &&  + 
\left. \textstyle \frac{1}{2}c_2|\Theta|^2 +\sum_{j=-2}^{+2}\phi_{j}^*\Gamma_{j} \right],
\end{eqnarray}
are two conserved quantities of an 
SO-coupled BEC. In the presence of SO-coupling ($\gamma \ne 0$), magnetization ($\equiv \int F_z dx dy= \int dxdy[2\rho_{+2}(x,y)-2\rho_{-2}(x,y)+\rho_{+1}(x,y)-\rho_{-1}(x,y)]$) is not a 
conserved quantity, although it is conserved for $\gamma=0$.

\subsection{Phase Requirement}
\label{section2a}
The permitted vortex configurations in a spinor BEC depend on the 
inter-component phase relationships. Considering a circular symmetry, the 
spinor order parameter for a vortex configuration in circular polar 
co-ordinates $(r,\theta)$ can be written in terms of amplitude and phase 
part as
\begin{equation}
\phi_j(r,\theta) = R_j(r) e^{i(w_j\theta+\alpha_j)},
\label{anstaz}
\end{equation}
where $R_j = |\phi_j(r,\theta)| \geq 0$ and $j =0,\pm 1,\pm 2$. The phases 
of the component wave functions have contributions from winding number 
$w_j$ of the phase-singularity which is an integer and any other constant 
phase $\alpha_j$. Using the {\em ansatz} (\ref{anstaz}), one can 
minimize the interaction and the energy contribution from the 
SO coupling leading to following independent relationships among the permitted 
winding numbers (details are given in Appendix):
 \begin{subequations}\label{windingrelations}
\begin{align}
 w_{+2}-w_{+1}+1 &= 0,
\quad w_{+1}-w_0+1=0, \\
w_{-2}-w_{-1}-1&=0,
\quad w_{-1}-w_0-1=0.
\end{align}
\end{subequations}
The allowed winding-number combinations are $(-2,-1,0,+1,+2)$, $(-1,0,+1,+2,+3)$, 
$(0,+1,+2,+3,+4)$, and higher. It is to be also noted that {an} 
axisymmetric configuration without any phase singularity in any 
of the components, i.e., with a winding number combination of (0,0,0,0,0), 
is not allowed as per 
Eqs.~(\ref{windingrelations}a)-(\ref{windingrelations}b). Using 
Eq.~(\ref{anstaz}), kinetic energy (KE) of the condensate is
\begin{equation}
{\rm KE} =\sum_{j=-2}^2 w_j^2\int \frac{\pi \phi_j^2}{r}dr,
\label{ke}
\end{equation}
which indicates that the system might end up favoring small winding numbers.

\subsection{Single-Particle Hamiltonian}
\label{section2b}

\begin{figure}[t]
\begin{center}
\includegraphics[trim = 0cm 0cm 0cm 0cm, clip,width=\linewidth,clip]{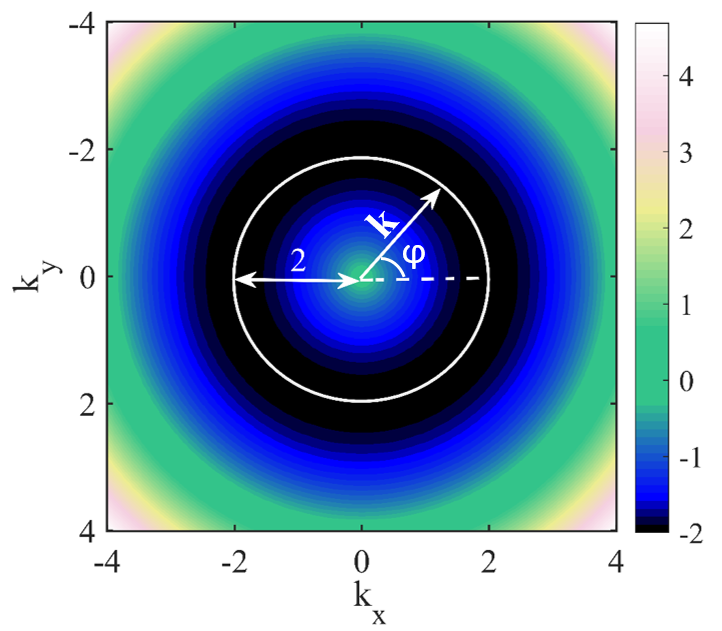}
\caption{(Color online)  Contour plot of eigen energy $E(k_x,k_y)$ in 
Eq.~(\ref{minima}) for $\gamma = 1$. The minima corresponding to $k_x^2+k_y^2 = 4$ is
a circle of radius $2$. A typical ${\bf k}$ with magnitude $2$ and oriented at a 
polar angle $\varphi$ is also shown. }
\label{minima-ring}
\end{center}
\end{figure}

The emergence of the axisymmetric solutions to Eqs. 
(\ref{cgpet3d-1})-(\ref{cgpet3d-3}) in the form of a $(-2,-1,0,+1,+2)$-type multi-ring state
can be inferred from the eigenfunction of the single-particle (or non-interacting) Hamiltonian in 
Eq.~(\ref{sph}). {\it One} eigenfunction  of the single-particle Hamiltonian with 
(minimum) energy 
 $-2\gamma^2$ is 
\begin{equation}
\Phi = \frac{1}{4}\begin{pmatrix}
e^{-2i\varphi} \\
-2 e^{-i\varphi}\\
\sqrt{6} \\
-2 e^{i\varphi}\\
e^{2i\varphi}\\
\end{pmatrix}e^{ixk_x+iyk_y} \equiv \zeta(\varphi)e^{ixk_x+iyk_y}, 
                                  \label{ef_sph}
\end{equation}
where $\varphi$ = $\tan^{-1}(k_{y}/k_{x})$ and $k^2= 
{k_{x}^2+k_{y}^2} = (2\gamma)^2$ {which corresponds 
to the minimum of  eigen energy 
\begin{equation}
E(k_x,k_y) = \frac{1}{2}\left(k_x^2+k_y^2-4\gamma\sqrt{k_x^2+k_y^2}\right).
\label{minima}
\end{equation}
The two-dimensional contour plot of eigen energy $E(k_x,k_y)$ for $\gamma = 1$ 
is shown in Fig.~\ref{minima-ring}. The eigen energy is minimum along a circle
of radius $2$, i.e., for $k_x^2 + k_y^2 = 4$. Hence a typical ${\bf k} \equiv (k_x,k_y)$ 
which minimizes the eigen energy is as shown in Fig.~{\ref{minima-ring}}, where 
$\varphi$ can vary from $0$ to $2\pi$.}
The eigenfunctions with different 
orientations of the vector ${\bf k}\equiv\{k_x,k_y\}$ in the {$k_x-k_y$ plane,
as shown in Fig.~\ref{minima-ring}}, are all degenerate. Thus
a most general solution to the single-particle Hamiltonian can be obtained 
by considering the superposition of eigenfunctions (\ref{ef_sph}) 
with ${\bf k}$ allowed to point along all directions in 2D plane. The 
generic solution, so obtained, is 
\begin{align}\label{abcd}
\Phi_{\rm MR} &= \frac{1}{8\pi}\int_{0}^{2 \pi}\begin{pmatrix}
e^{-2i\varphi} \\
-2 e^{-i\varphi}\\
\sqrt{6} \\
-2 e^{i\varphi}\\
e^{i2\varphi}\\
\end{pmatrix}e^{i2\gamma r\cos(\varphi-\theta)}d\varphi,\\
&= \frac{1}{4}\begin{pmatrix}
-e^{-2 i \theta } J_2( 2 \gamma r)\\
-2i e^{- i \theta } J_1( 2 \gamma r)\\
\sqrt{6}J_0( 2 \gamma r)\\
 -2i e^{ i \theta } J_1(2 \gamma r)\\
-e^{2 i \theta } J_2( 2 \gamma r)\label{gen_ef_sph}
\end{pmatrix},
\end{align}
where $\theta = \tan^{-1}{y/x}$, and $J_n(2 \gamma r)$ with $n=0,1,2$ is 
the Bessel function of first kind of order $n$ and where $\Phi_{\rm MR} $ has the 
phase singularities of a multi-ring (MR) soliton. Solution 
(\ref{gen_ef_sph}) agrees with the permissible winding number 
combination of $(-2,-1,0,+1,+2)$ obtained earlier based on energetic
considerations, viz. Eq.(\ref{windingrelations}), and corresponds to a 
$(-2,-1,0,+1,+2)$-type multi-ring soliton. As the component densities, $\rho_j \sim |J_{|j|}|^2$, the 
densities would have a long  undulating tail, and in the asymptotic region 
with $r\rightarrow\infty$, 
$\rho_j\sim \sqrt{2/(2\pi \gamma r)}\cos\left( 2 \gamma r-\pi |j|/2-\pi/4\right)$.

\begin{figure}[t]
\begin{center}
\includegraphics[trim = 0cm 0cm 0cm 0cm, clip,width=\linewidth,clip]{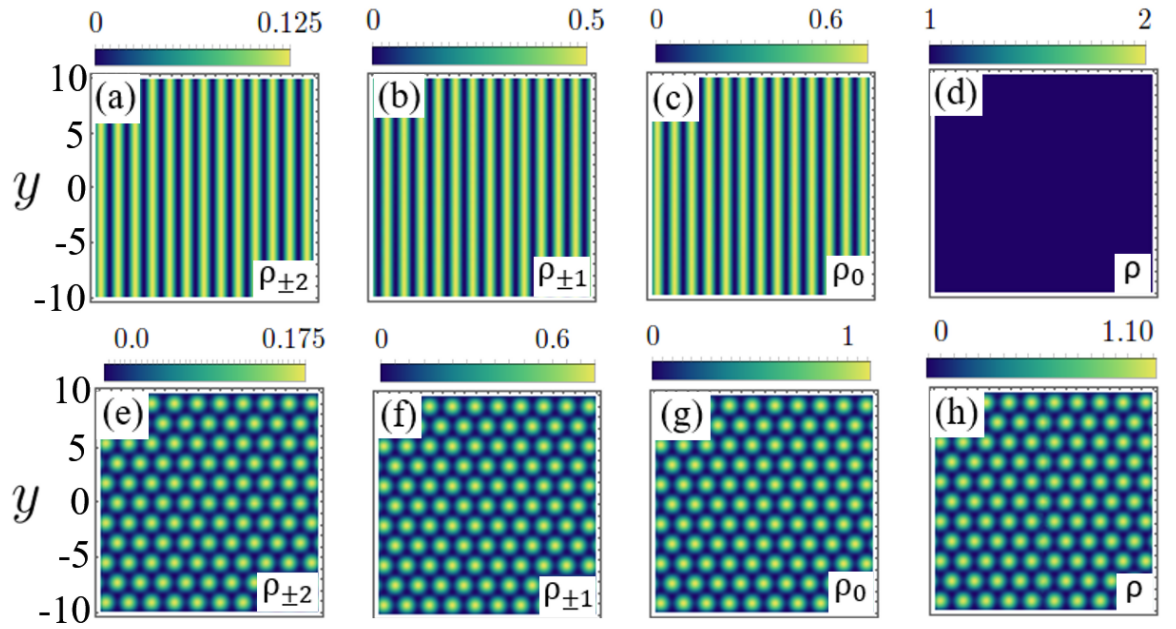}
\includegraphics[trim = 0cm 0cm 0cm 0cm, clip,width=\linewidth,clip]{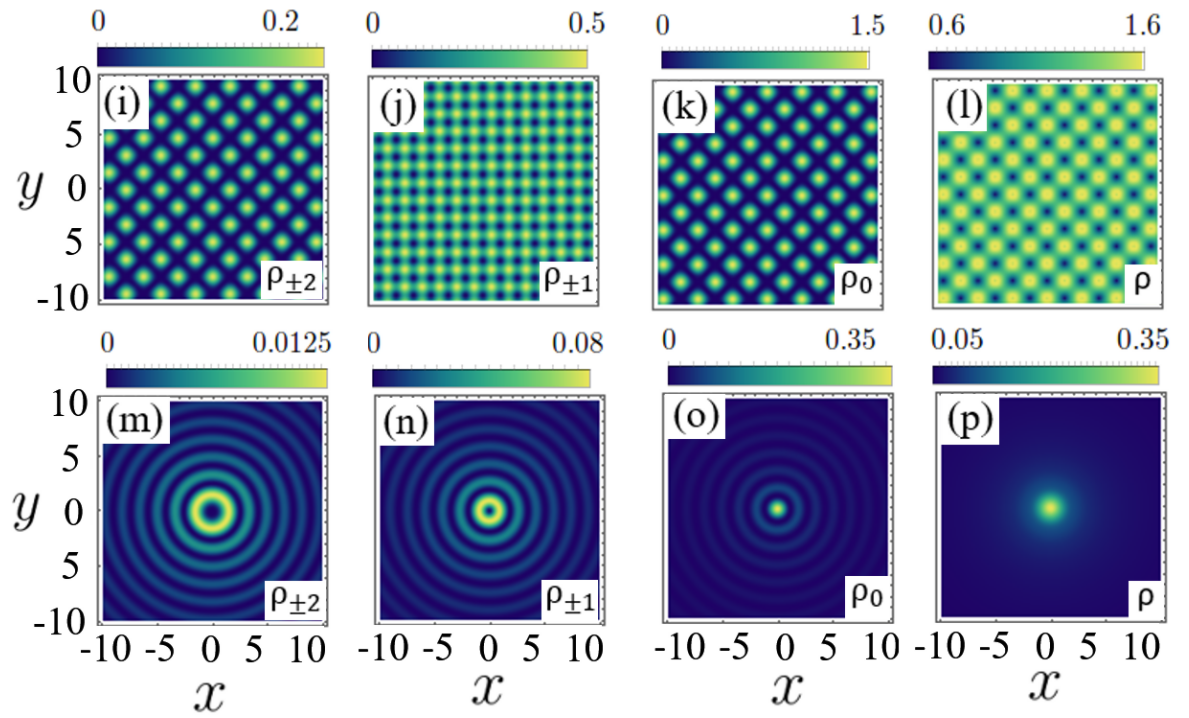}
\caption{(Color online) {The two-dimensional contour plot of densities of 
the components $j = \pm2$, $j = \pm1$, $j = 0$, and total density corresponding to 
$\Phi_{\rm ST}$ is shown in (a)-(d) for SO-coupling strength $\gamma =1$. The same for
$\Phi_{\rm TL}$, $\Phi_{\rm SL}$ and $\Phi_{\rm MR}$ are shown in (e)-(h), (i)-(l) and (m)-(p), respectively. }}
\label{single-particle-analysis}
\end{center}
\end{figure}

Besides superposition of an infinite number of plane waves, viz. Eq. \ref{abcd},
one can also have a superposition of (a) two counter-propagating plane waves, (b) three plane 
waves whose propagation vectors make an angle $2\pi/3$ with each other, or 
(c) four plane waves whose propagation vectors make an angle $\pi/2$ with each 
other. Choosing $x$-direction as the direction for one of these wave vectors, 
these superpositions, representing a stripe (ST), triangular lattice (TL),
and square lattice (SL), respectively, are
\begin{subequations}\label{phi_sl}
 \begin{align}
  \Phi_{\rm ST} &= \textstyle \frac{1}{\sqrt{2}}\left[\zeta(0)e^{i2\gamma x}
                 + \zeta(\pi)e^{-i2\gamma x}\right],\\
  \Phi_{\rm TL} &=\textstyle \frac{1}{\sqrt{3}} \left[\zeta(0)e^{i2\gamma x}
                 + \zeta(2\pi/3)e^{i(-\gamma x+\gamma\sqrt{3}y)}\right.
                   \nonumber\\
                &\left.+\zeta(4\pi/3)e^{i(-\gamma x-\gamma\sqrt{3})}\right],\\
  \Phi_{\rm SL} &= \textstyle \frac{1}{2}\left[\zeta(0)e^{i2\gamma x}
                +\zeta(\pi/2)e^{i2\gamma y}+\zeta(\pi)e^{-i2\gamma x}\right.
                \nonumber\\
                &\left.+\zeta(3\pi/2)e^{-i2\gamma y}\right].
 \end{align}
\end{subequations}
{ The component densities and corresponding total density for these degenerate 
solutions corresponding to $|\Phi_{\rm ST}|^2$, $|\Phi_{\rm TL}|^2$, $|\Phi_{\rm SL}|^2$ 
and $|\Phi_{\rm MR}|^2$, are shown in Figs. \ref{single-particle-analysis}(a)-(d), 
\ref{single-particle-analysis}(e)-(h), \ref{single-particle-analysis}(i)-(l), and
\ref{single-particle-analysis}(m)-(p), respectively.}
If one examines the total density corresponding to these superpositions in 
Eqs.~(\ref{gen_ef_sph})-(\ref{phi_sl}) {as plotted in 
Figs.~{\ref{single-particle-analysis}}(d), (h), (l), and (p)}, then in the total density corresponding 
to $|\Phi_{\rm MR}|^2$ and $|\Phi_{\rm ST}|^2$ there is no spatially-periodic modulation,  
whereas the total density $|\Phi_{\rm TL}|^2$ and $|\Phi_{\rm SL}|^2$ will have a hexagonal and a 
square-lattice crystallization, respectively. The localized solitons studied in this paper 
can be qualitatively approximated by the single-particle solutions  (\ref{gen_ef_sph})-(\ref{phi_sl})
multiplied by a localized Gaussian function.  In the numerical solution by an imaginary-time propagation
such approximations can be used as the initial functions for different solitons with appropriate symmetry. 

%%%%%%%%%%%%%%%%%%%%%%%%%%%%%%%%%%%%%%%%%%%%%%%%%%%%%%%%%%%%%%%%%%%%%%%%%%%%%%%%
%%%%%%%%%%%%%%%%              Numerical Results
%%%%%%%%%%%%%%%%%%%%%%%%%%%%%%%%%%%%%%%%%%%%%%%%%%%%%%%%%%%%%%%%%%%%%%%%%%%%%%%%

\section{Numerical results}
\label{section3}

We numerically solve the GP equations  (\ref{cgpet3d-1})-(\ref{cgpet3d-3}) using split 
time-step Fourier spectral method \cite{Paramjeet}. For SOC strengths upto 
$\gamma$ = 1, we consider the  spatial  step sizes 
$\Delta x = \Delta y = 0.1$. Here the 
two-dimensional box size for solving the GP equations  is $60\times60$. For $\gamma>1$, 
 the step sizes and box size considered are $\Delta x = \Delta y = 0.05$, 
$\Delta t= 0.00025$ and $40\times40$, respectively. The time steps for imaginary- 
and real-time propagation are $\Delta t=0.1\times \Delta x^2 $ and $\Delta t=0.05\times \Delta x^2 $,
respectively. The imaginary-time propagation method is used for finding the lowest-energy state of
a specific symmetry, whereas real-time propagation is used to study the dynamics.
The initial guess for order 
parameter to obtain the stripe, triangular-lattice and 
square-lattice solitons are considered as solutions to non-interacting condensate, 
viz. Eqs.~(\ref{phi_sl}a)-(\ref{phi_sl}c), multiplied by a localized Gaussian state
and the same for the multi-ring soliton 
is a two-dimensional Gaussian function with appropriate vortices phase-imprinted on different components. 
As magnetization is not conserved, during time propagation magnetization 
is allowed to evolve freely and attain a final converged value independent of the magnetization 
of the initial state.
The dynamic stability of the solutions is demonstrated by real-time evolution with
a small random noise added to the order parameter at $t=0$, wherein they 
retain their structure over long periods of evolution.

\begin{figure}[t]
\begin{center}
\includegraphics[trim = 0cm 0cm 0cm 0cm, clip,width=\linewidth,clip]{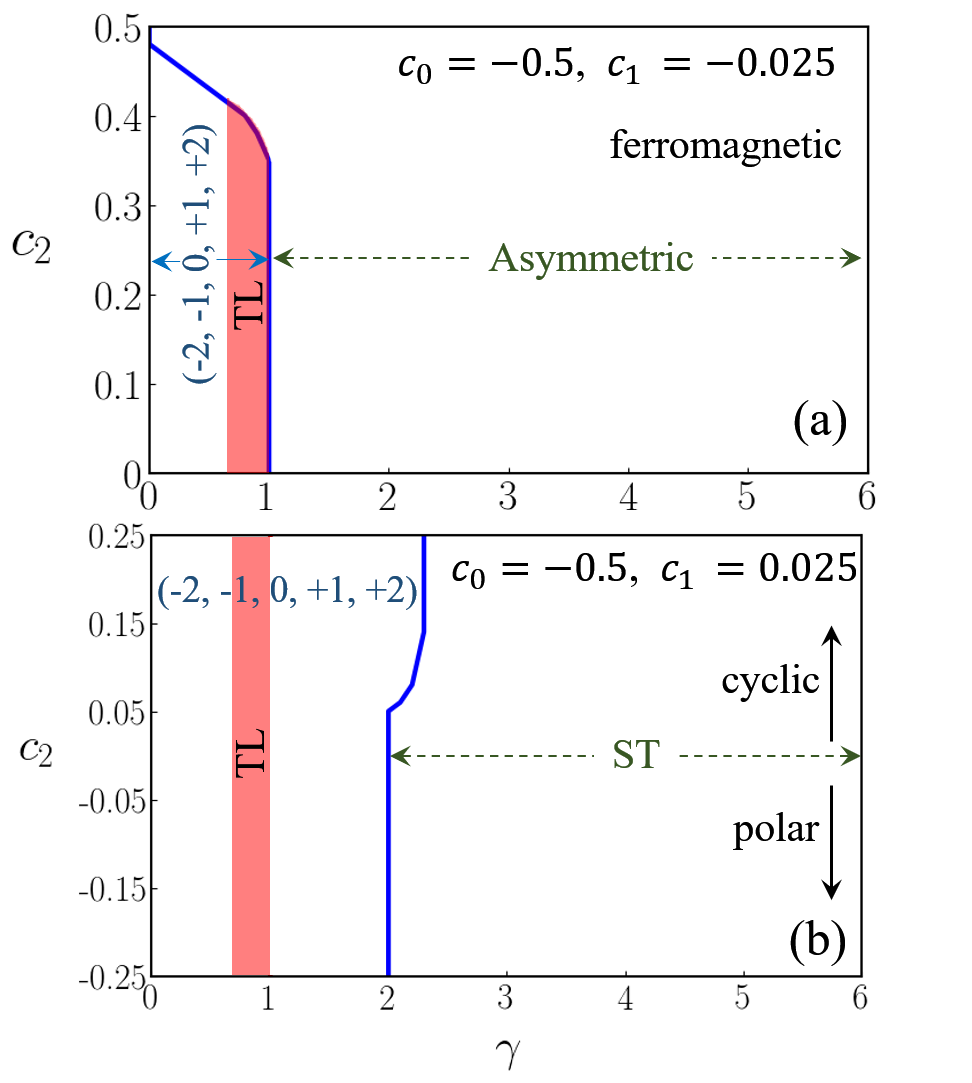}
\caption{(Color online) The $c_2$ versus $\gamma$ phase plots for the ground states
are shown (a) for the ferromagnetic phase with $c_0=-0.5$ and $c_1=-0.025$ and (b) for the
cyclic and polar phases with $c_0=-0.5$ and $c_1=0.025$. In (a) for small SO-coupling strengths, 
the axisymmetric $(-2,-1,0,+1,+2)$ state is the ground state similar to the state shown
in \ref{fig1}(a)-(d), whereas for larger SO-coupling strengths, the asymmetric soliton state is the ground 
state similar to the state shown in \ref{fig2}(f)-(j). In (b) for small SO-coupling strengths, the 
axisymmetric $(-2,-1,0,+1,+2)$ state is the ground state similar to the state shown in \ref{fig1}(a)-(d), 
whereas for larger SO-coupling strengths, the stripe soliton state (ST) is the ground state similar to the
state shown in \ref{fig5}(a)-(d). Difference in energy of other quasi-degenerate solitons 
from these ground states is $\gtrapprox 10^{-4}$. As an illustration, one of the quasi-degenerate states is
the triangular-lattice (TL) state similar to the
state shown in \ref{fig3}(a)-(d), which occupies a narrow region near $\gamma=1$ and is shown by a red 
shaded strip. The energy differences between the triangular lattice (TL) state  and the $(-2,-1,0,+1,+2)$-type
multi-ring ground state in plots (a) and (b) are $10^{-4}$-$10^{-3}$.}
\label{fig1x}
\end{center}
\end{figure}

Our numerical studies 
reveal that an SO-coupled spin-2 BEC with attractive interactions can have a 
variety of self-trapped stationary solutions including the cases where the total 
density of the condensate exhibits regular hexagonal or square patterns.
The ground-state phase diagram of a ferromagnetic BEC with $c_0 = -0.5$, $c_1 = -0.025$ 
and polar and cyclic BECs with $c_0 = -0.5$, $c_1 = 0.025$  in $c_2$-$\gamma$ planes are shown in Figs. 
\ref{fig1x}(a) and (b), respectively. For the ferromagnetic BEC, as the strength of the SO coupling is increased, the ground state changes from  an 
 axisymmetric multi-ring soliton to   an asymmetric
soliton. For the polar and cyclic BECs, the ground-state 
phase changes from a multi-ring soliton to a stripe soliton above a critical SO-coupling for the chosen set of 
interaction parameters. In a narrow strip near $\gamma = 1$, a triangular-lattice soliton appears as one of 
the quasi-degenerate ground states in all three magnetic phases as shown by shaded regions in 
Figs. \ref{fig1x}(a) and (b).  
It is also pertinent to point out that the energy difference among the 
quasi-degenerate states decreases  (increases)  with a decrease (increase)  in $|c_0|$.

%The energy of these various solitons are different at the 3rd decimal place 
%following that these are characterized as ground state phases in various domains.
%In the cyclic and polar phase, for all values of SO-coupling strength $\gamma$, 
%the lowest energy state is a stripe phase. In addition, for small SO-coupling 
%strengths, (-2,-1,0,+1,+2) and hexagonal lattice also appear as the minimum
%energy solution. For the intermediate strength of SO-coupling, (-1,0,+1,+2,+3)
%appears as the ground state solution in addition to the stripe phase.\\
%For the ferromagnetic domain, for all values of SO-coupling strength $\gamma$,
%the lowest energy state is an asymmetric phase. In addition, for small 
%SO-coupling strengths, (-2,-1,0,+1,+2), (-1, 0,+1,+2,+3) and hexagonal 
%lattice also appear as the minimum energy solution. For the intermediate 
%strength of SO-coupling, stripe phase appears as the ground state solution 
%in addition to the asymmetric phase. 

\subsection{Small SO-coupling strength}\label{section3a}
\subsubsection{Ferromagnetic Phase}

\begin{figure}[t]
\begin{center}
\includegraphics[trim = 0cm 0cm 0cm 0cm, clip,width=\linewidth,clip]{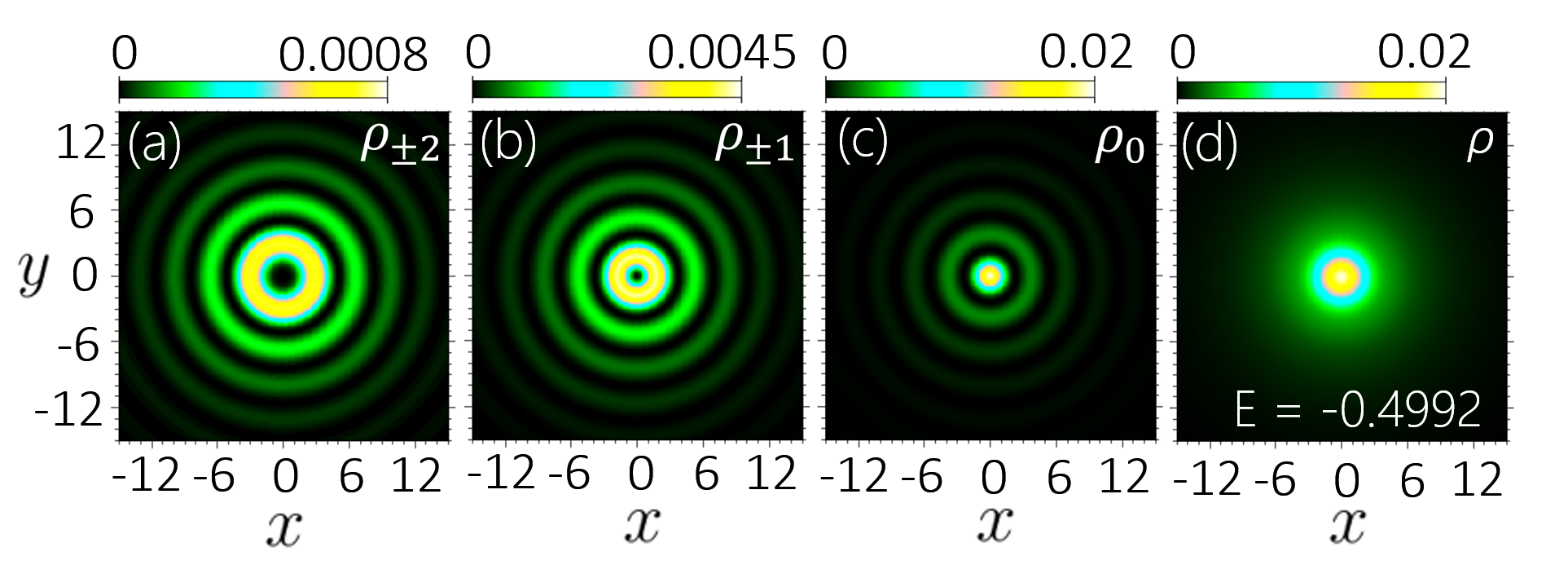}
\caption{(Color online) Contour plot of density of the components (a) $j = \pm 2$, (b)  $j = \pm1$, 
(c) $j = 0$, and (d) total density of an axisymmetric  $(-2, -1, 0, +1, +2)$-type  multi-ring  soliton   
with $c_0 = -0.5$, $c_1 = -0.025$ ,  $c_2 = 0.25$, (ferromagnetic phase) and 
$\gamma =0.5$ with energy  $E= -0.4992$. }
\label{fig1}
\end{center}
\end{figure}

In an SO-coupled spin-2 BEC with $c_0<0$, $c_1^{(1)}\leq c_1 <0$, and $c_2>0$, 
which implies that the system is weakly ferromagnetic, {where $ c_1^{(1)}$ is a constant, viz. 
Fig. \ref{phase},} the lowest-energy state has an axisymmetric density pattern corresponding to a 
$(-2,-1,0,+1,+2)$-type  multi-ring 
soliton, whereas the higher energy states could be axisymmetric or circularly asymmetric. For smaller  
$c_1$, i.e. $c_1<c_1^{(1)}$, the interactions become (relatively) strongly ferromagnetic, and the ground 
state corresponds to a circularly-asymmetric soliton. The $(-2,-1,0,+1,+2)$-type axisymmetric multi-ring 
soliton continues to exist in this case, but is no longer the ground 
state. With further decrease of $c_1$ below another constant $c_1^{(2)}$ an increase of 
attractive interaction leads to a collapse of the condensate and no solution 
exists. The explicit values of the constants $c_1^{(1)}$ and $c_1^{(2)}$ are dependent on 
the parameters $c_0$, $c_2$ and $\gamma$.

\begin{figure}[t]
\begin{center}
\includegraphics[trim = 0cm 0cm 0cm 0cm, clip,width=\linewidth,clip]
               {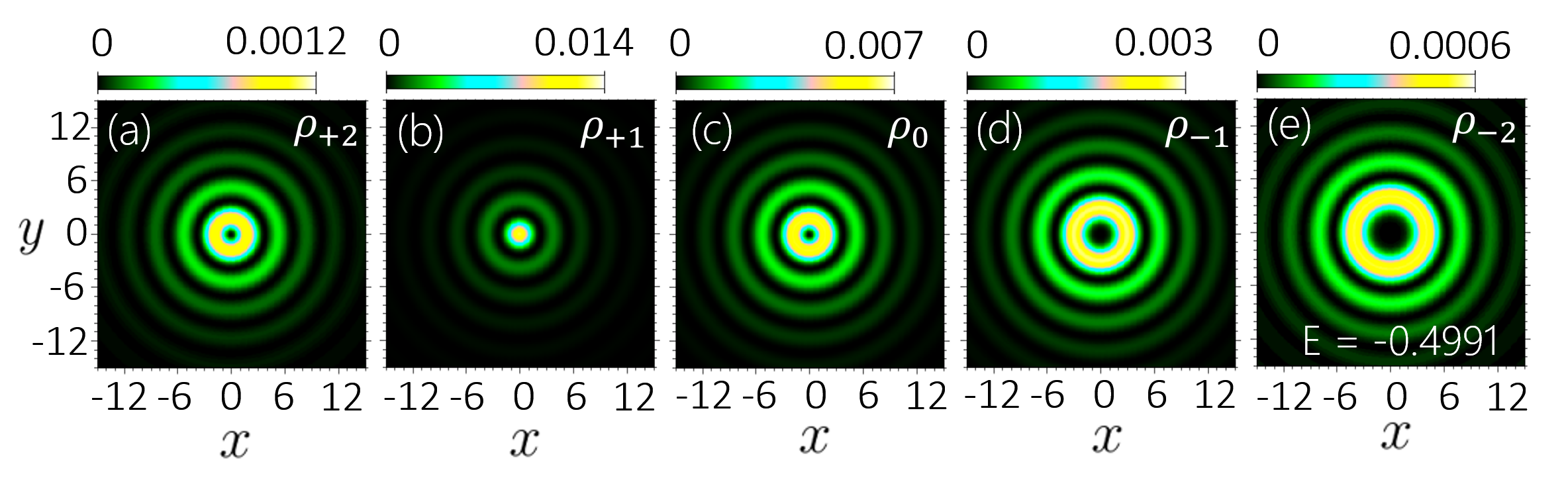} 
 \includegraphics[trim = 0cm 0cm 0cm 0cm, clip,width=\linewidth,clip]
               {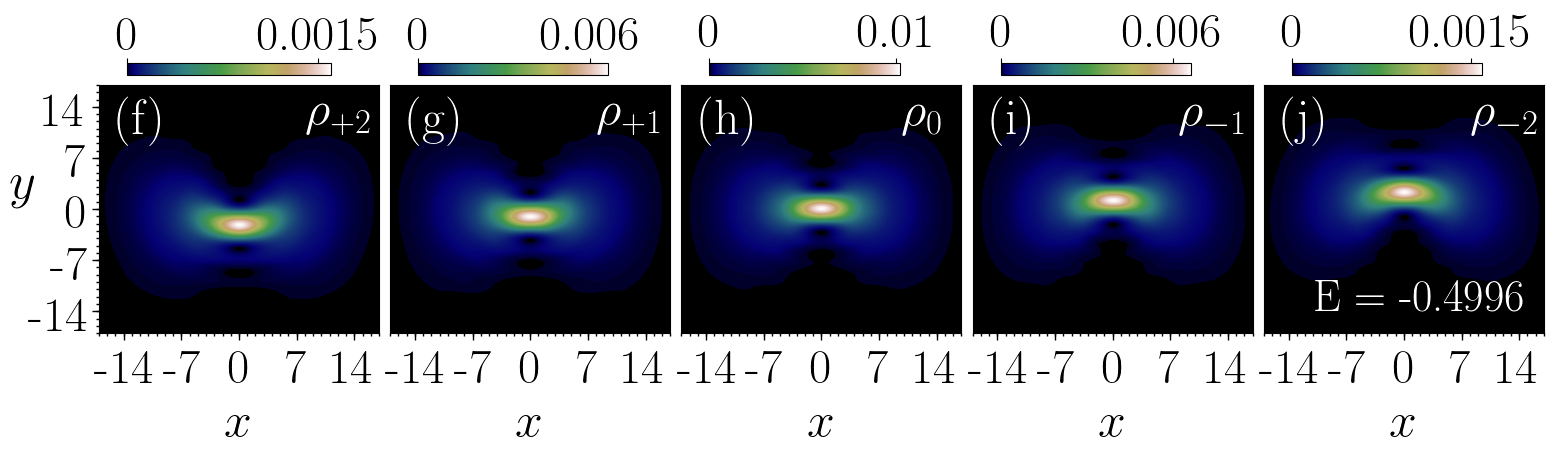}              
\caption{(Color online)  Contour plot of density of 
components (a) $j = +2$, (b) $j = +1$,  (c) $j = 0$, (d)  $j = -1$, and (e) $j = -2$ of an 
axisymmetric  $(-1, 0, +1, +2, +3)$-type multi-ring  soliton with 
$c_0 = -0.5$, $c_1 = -0.025$, $c_2 = 0.25$, 
(ferromagnetic phase) $\gamma =0.5$  and energy $E= -0.4991$;
the same of a circularly-asymmetric soliton with $c_0 = -0.5$ $c_1 = -0.1$, 
$c_2 = 0.25$, (strongly ferromagnetic phase),  $\gamma =0.5$, and energy $E = -0.4996$
in (f)-(j)..} 
\label{fig2}
\end{center}
\end{figure}

{\em Axisymmetric multi-ring soliton}: As an example in the ferromagnetic 
phase, we consider $c_0 = -0.5$, $c_1 = -0.025 > c_1^{(1)}=-0.05$, 
$c_2 = 0.25$ and $\gamma = 0.5$. The ground state solution for this set of 
parameters is an axisymmetric $(-2, -1, 0, +1, +2)$-type multi-ring 
soliton with energy $E = -0.4992$ as 
exhibited in Fig.~({\ref{fig1}}) through a contour
plot of the component densities (a) $\rho_{\pm 2}$, (b) $\rho_{\pm 1},$ (c) $\rho_0$
and (d) the total density $\rho$. The densities of components $\pm j$ with $j=1,2$ 
are equal. This state  has the same rotational symmetry as  the ground state of the non-interacting 
SO-coupled condensate governed  by Eq.~(\ref{gen_ef_sph}) and  has a long undulating
tail of decreasing amplitude consistent with the asymptotic behaviour of Bessel 
functions.  If the wave function  (\ref{gen_ef_sph}) is multiplied by a localized Gaussian function,
the resultant function qualitatively produces the density of the state displayed in Fig. \ref{fig1}.
Hence the density and symmetry properties of  the actual physical state can be inferred from a study 
of the eigenfunctions of the single-particle Hamiltonian.
The total density has no core at the center as the vortex cores  of $j = \pm 2$, and 
$j = \pm 1$ components  are filled by a {non-zero density} at the center of the $j = 0$ component. 
The first zeros of $J_0(r), J_1(r)$, and $J_2(r)$ are $2.40483$, $3.83171$, and $5.13562$, 
respectively, and these agree very well with the results presented in 
Fig.~(\ref{fig1}). %{\color{red}, where most of the condensate is accommodated 
%within region below first zeros, and thus obscures the visibility of the 
%rings.}
Numerically, this solution is obtained by evolving Eqs.  
(\ref{cgpet3d-1})-(\ref{cgpet3d-3}) in imaginary time and using, as an initial 
guess, a two-dimensional Gaussian function multiplied by an appropriate phase 
factor of $\exp(-ij\varphi)$ for the $j$th component. For the same set of 
parameters, we also have a $(-1, 0, +1, +2, +3)$-type  multi-ring 
soliton with an energy -0.4991 as shown in Fig.~\ref{fig2}(a)-(e) through a contour plot of component densities 
(a) $\rho_{+2}$, (b) $\rho_{+1}$, (c) $\rho_{0}$, (d) $\rho_{-1}$, and (e) $\rho_{-2}$.
The quasi-degeneracy between the two solutions shown in Figs. \ref{fig1}(a)-(d) and \ref{fig2}(a)-(e) 
is lifted with an increase in $|c_0|$. The winding number combinations 
for these axisymmetric solutions are in accordance with relations 
given in Eq.~(\ref{windingrelations}). %Lower winding number combination 
%$(-2, -1, 0, +1, +2)$ appears as ground state as it minimizes the kinetic 
%energy as per  %Eq.~(\ref{windingrelations}). 
%Eq. (\ref{ke}).

{\em Circularly-asymmetric soliton}: 
By considering the parameters in the (relatively) strongly ferromagnetic 
phase with $c_0 = -0.5$, $c_1 = -0.1$, $c_2 = 0.25$ and $\gamma = 0.5$, 
the  circularly-asymmetric soliton displayed in Figs.~{\ref{fig2}}(f)-(j) 
through a contour plot of component densities turns out to be the 
ground state while the axisymmetric $(-2, -1, 0, +1, +2)$-type multi-ring soliton appears 
as an excited state. Asymmetry  of the solution arises, in this case, as
different from the $(-2, -1, 0, +1, +2)$-type soliton displayed in Figs. \ref{fig1}(a)-(d),
the phase-singularities in $\pm j$ 
components of the circularly-asymmetric soliton exhibited in Figs.~{\ref{fig2}}(f)-(j)
do not overlap. When we keep on decreasing $c_1$ further, then 
these singularities in $\pm j$  components  move further apart along $y$-axis. 
For $c_0 = -0.5$, $c_1 = -1.3$, $c_2 = 0.25$ and $\gamma = 0.5$, 
phase-singularities lie in the region where the condensate density is quite 
small (not shown here) and hence no perceptible density hole is visible in 
the component densities
$\rho_j$. If we decrease $c_1$ below $c_1^{(2)} = -1.3$, while keeping 
$c_0$, $c_2$, and $\gamma$ fixed, then the condensate collapses.

\subsubsection{Cyclic and Polar phases}

\begin{figure}[t]
\begin{center}
\includegraphics[trim = 0cm 0cm 0cm 0cm, clip,width=1\linewidth,clip]{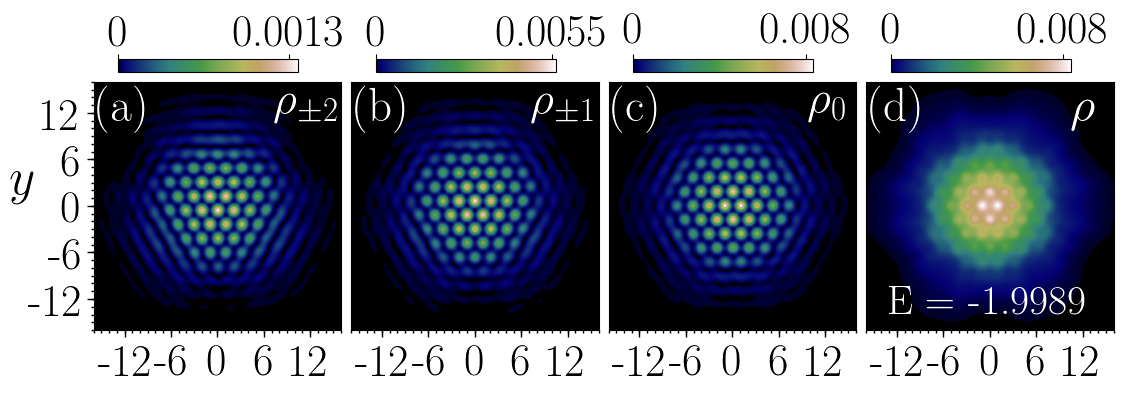}
\includegraphics[trim = 0cm 0cm 0cm 0cm, clip,width=1.05\linewidth,clip]{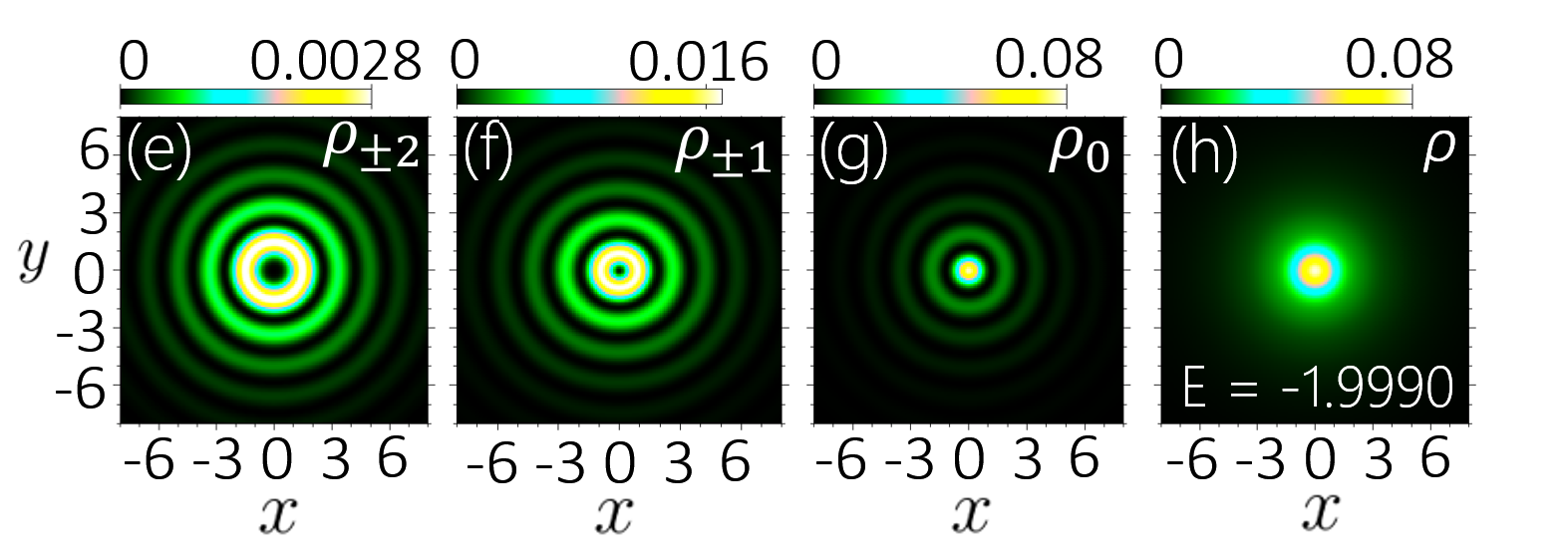}

\caption {(Color online) Contour plot of density of components 
(a) $j = \pm{2}$, (b) $j = \pm{1}$, (c) $j = 0$, and  (d) total density of  a triangular-lattice 
soliton with $c_0 = -0.5$, $c_1 = 0.025$, $c_2 = 0.25$, (cyclic phase)  $\gamma =1$, and  energy $E = -1.9989$; 
the same of a $(-2,-1,0,+1,+2)$-type  multi-ring soliton for the same parameters and  $E = -1.9990$  in (e)-(h). }
\label{fig3}
\end{center}
\end{figure}

For small SO-coupling strengths, in both cyclic and polar phases, similar to the ferromagnetic phase, 
the axisymmetric $(-2, -1, 0, +1, +2)$-type multi-ring soliton emerges as the ground 
state, whereas the axisymmetric $(-1, 0, +1, +2, +3)$-type  soliton appears as a metastable 
state (result not shown  here). For example, with $c_0 = -0.5$, $c_1 = 0.025$, $c_2 = 0.25$ and 
$\gamma = 0.5$ corresponding to the cyclic phase, viz. Fig. \ref{phase}, the axisymmetric $(-2,-1,0,+1,+2)$-type 
and $(-1,0,+1,+2,+3)$-type multi-ring solitons have energies $-0.4992$ and $-0.3864$, 
respectively. Similarly, with  $c_0 = -0.5$, $c_1 = 0.025$, $c_2 = -0.25$ and 
$\gamma = 0.5$, corresponding to the polar phase, the respective energies of 
axisymmetric $(-2,-1,0,+1,+2)$-type and $(-1,0,+1,+2,+3)$-type  multi-ring solitons 
are $-0.4994$ and $-0.3855$. In both cases, the $(-2, -1, 0, +1, +2)$-type multi-ring 
soliton is the ground state.

\begin{figure}[t]
\begin{center}
\includegraphics[trim = 0cm 0cm 0cm 0cm, clip,width=1\linewidth,clip]
                {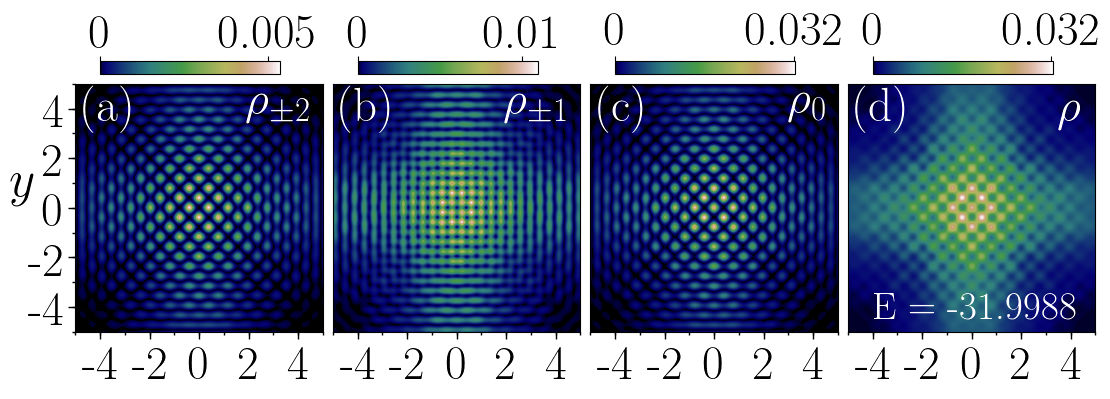}
\includegraphics[trim = 0cm 0cm 0cm 0cm, clip,width=1.02\linewidth,clip]
                {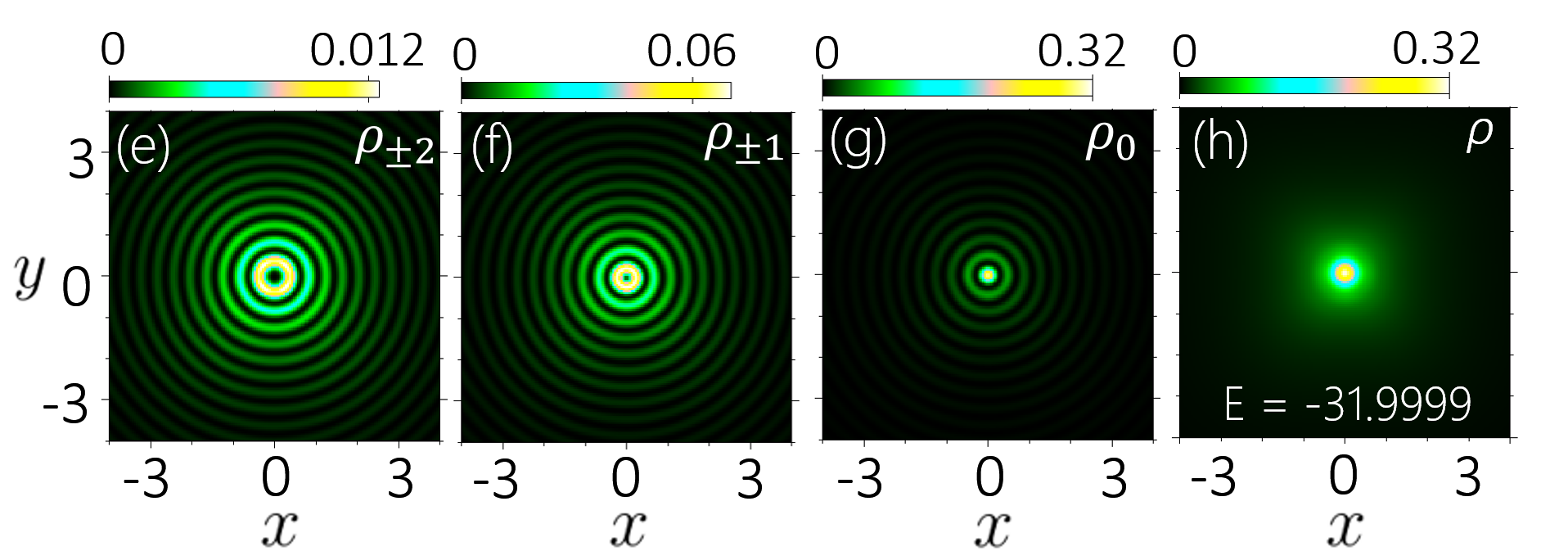}
\caption {(Color online)  Contour plot of density of a square-lattice soliton of  components 
(a) $j = \pm2$, (b) $j = \pm1$, (c) $j = 0$ and  (d) total density  for 
$c_0 = -0.5$, $c_1 = -0.025$, $c_2 = 0.25$, (ferromagnetic phase)
$\gamma = 4$ and $E=-31.9988$;  for the same parameters, the component and total densities of a
$(-2,-1,-0,+1,+2)$-type  multi-ring  soliton in (e)-(h) with $E=-31.9999$.}
\label{fig4}
\end{center}
\end{figure}

\subsection{Intermediate SO-coupling strength}
\label{section3b}

For  intermediate SO-coupling strengths, we get a 
triangular-lattice soliton, with a hexagonal-lattice crystallization in components and total densities,
in all three magnetic phases $-$ ferromagnetic, polar and cyclic. Although a square-lattice soliton
has been earlier identified in Ref.~\cite{adhikari2021multiring}, 
a triangular-lattice soliton  was not found  in the spin-1 
case.  For example, in the cyclic phase with 
$c_0 = -0.5$, $c_1 = 0.025$, $c_2 = 0.25$ and $\gamma = 1$, the  
triangular-lattice soliton  is shown in Fig.~\ref{fig3} through a contour density plot of component 
densities (a) $\rho_{\pm 2}$, (b) $\rho_{\pm 1}$, (c) $\rho_{0}$ and (d) total density.
 The triangular-lattice structure is a result of superposition of 
three plane waves and corresponds to  a solution of the  non-interacting system given 
by Eq.~(\ref{phi_sl}b). However, to get a localized hexagonal structure as in Figs. 
\ref{fig3}(a)-(d), the function  (\ref{phi_sl}b) has to be multiplied by a localized Gaussian function.
For the same parameters, an axisymmetric $(-2,-1,0,+1,+2)$-type multi-ring soliton corresponding to the 
single-particle solution (\ref{gen_ef_sph}) is also a solution as 
illustrated in Figs.~\ref{fig3}(e)-(h). Both these states, the  multi-ring
 and the triangular-lattice solitons, have approximately the same numerical 
energy ($E=-1.9990$ and $E=-1.9989$) and are quasi-degenerate.
This degeneracy between the two solutions is removed with an increase in $|c_0|$ resulting
in the $(-2,-1,0,+1,+2)$-type multi-ring soliton as the ground state.

\subsection{Large SOC strength}
\label{section3c}

When $\gamma$ is increased further, different types  of degenerate 
states  appear with approximately the same energy in the three different magnetic phases. As an example, 
in the ferromagnetic phase with $c_0 = -0.5$, $c_1 = -0.025$, 
$c_2 = 0.25$, and $\gamma \gtrapprox 4$,  we get  the following five types of quasi-degenerate solitons: 
(1)  a {square-lattice} soliton, where as shown in Figs.~\ref{fig4}(a)-(d), the component as well as the 
total densities show square-lattice crystallization
consistent with the single-particle order parameter ~(\ref{phi_sl}c), 
(2)  a  {$(-2,-1,0,+1,+2)$-type multi-ring} soliton, corresponding to the single-particle order 
parameter (\ref{gen_ef_sph}), as shown in Figs.~\ref{fig4}(e)-(h), 
(3)  a  {circularly-asymmetric} soliton, (4) a {stripe} soliton with stripe 
{modulation in component densities} corresponding to the single-particle order parameter  
(\ref{phi_sl})(a), and (5)  a  {superstripe soliton} which has stripe 
patterns in component densities $\rho_{\pm 1}$ and square-lattice
 crystallization  in  component densities $\rho_{\pm 2}$ and 
$\rho_{0}$ and  also total density. The three latter 
solitons  are not shown here.   In case of the square-lattice soliton, viz. Fig. \ref{fig4}(a)-(d), 
the square-lattice crystallization in components $j=\pm 2$ and 0 are quite similar, whereas 
the square-lattice pattern in the components $j=\pm 1$ is different.  The lattice in components
$j=\pm 2$ and 0 makes an angle of $45\degree$ with the lattice in component $j=\pm 1$ 
and this is consistent with density pattern corresponding
to $\Phi_{\rm SL}$ as shown in Figs. {\ref{single-particle-analysis}}(i)-(l).
The prominent square lattice in total density has the same alignment as in components $j=\pm 2$ and 0. 
A similar square-lattice soliton was predicted in an SO-coupled spin-1 spinor BEC \cite{adhikari2021multiring}.
The densities of components $j=\pm 2$ and $0$ ($j=\pm 1$)  of the square-lattice soliton of 
Figs.~\ref{fig4}(a)-(d) are quite similar to the densities of components $j=\pm 1$ ($j=0$) 
of the same in an SO-coupled spin-1 spinor BEC \cite{adhikari2021multiring};  the total densities 
in the two  cases are also quite similar. The energies of these five different types of solitons are,
respectively, $-31.9988,-31.9999, -32.0071, 
-31.9998,$ and $-32.0020$ and hence these solitons are quasi-degenerate.

\begin{figure}[t]
\begin{center}
\includegraphics[trim = 0cm 0cm 0cm 0cm, clip,width=\linewidth,clip]
                {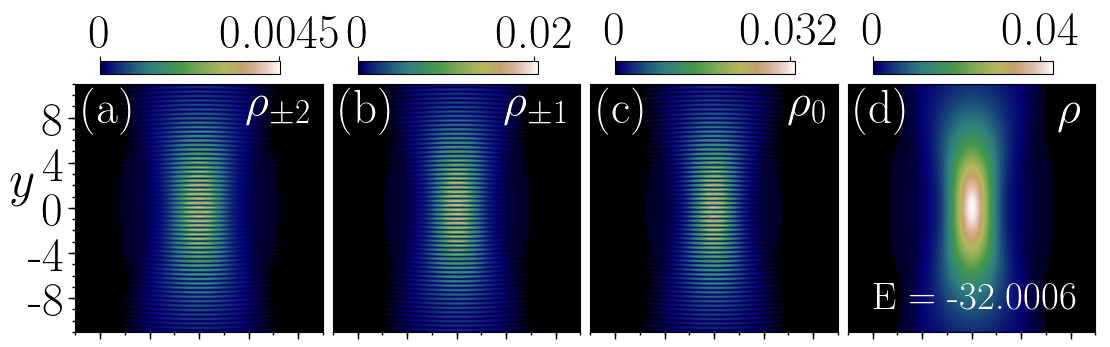}
\includegraphics[trim = 0cm 0cm 0cm 0cm, clip,width=\linewidth,clip]
                {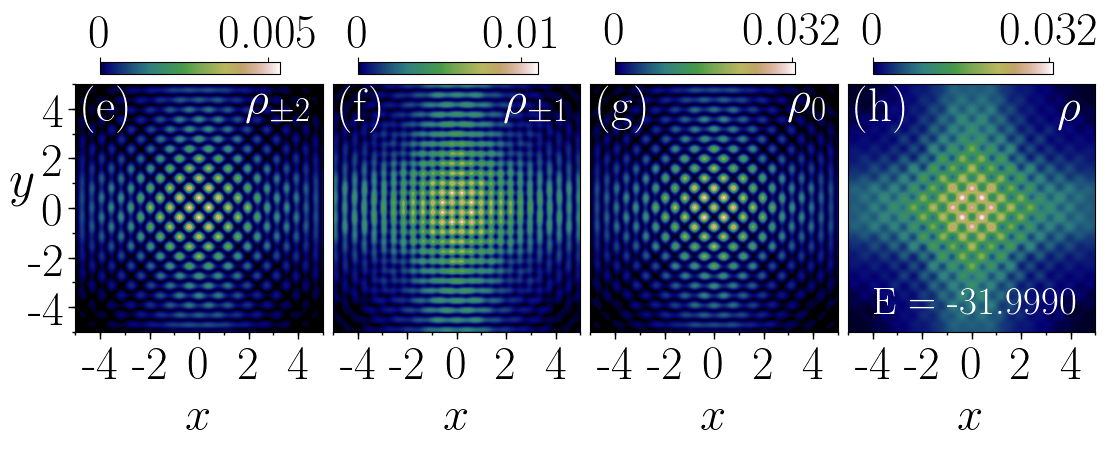}
\caption{(Color online) Contour plot of density of a stripe soliton of
components (a) $j = \pm 2$,  (b) $j = \pm 1$, (c) $j = 0$, and  (d) total density with 
$c_0 = -0.5$, $c_1 = 0.025$, $c_2 = 0.25$, (cyclic phase)   $\gamma = 4$, $E=-32.0006$; 
the same of a square-lattice soliton for the same parameters in (e)-(h) with 
$E=-31.9990$.
}
\label{fig5}
\end{center}
\end{figure}

In the cyclic phase, with $c_0 = -0.5$, $c_1 = 0.025$, $c_2 = 0.25$, 
and $\gamma \gtrapprox 4$, we again obtain four of the aforementioned 
quasi-degenerate solitons except the circularly-asymmetric soliton. The component and 
total densities corresponding to  {stripe} soliton and 
{ square-lattice} soliton are shown in Figs.~\ref{fig5}(a)-(d) and 
\ref{fig5}(e)-(h), respectively. The stripe soliton of Figs.~\ref{fig5}(a)-(d) is quite 
similar to the one in an SO-coupled spin-1 spinor BEC
\cite{adhikari2021multiring} for $\gamma=4$ in both ferromagnetic and  polar phases. 
In both cases the stripe pattern appears only in the component densities with the 
total density showing no modulation. The respective energies of the  stripe and the square-lattice solitons 
are $-32.0006$ and $-31.9990$.  The {superstripe soliton}  
with energy $-32.0002$ and  multi-ring soliton with energy $-31.9999$ are not shown here.

\begin{figure}[t]
\begin{center}
\includegraphics[trim = 0cm 0cm 0cm 0cm, clip,width=1\linewidth,clip]
                {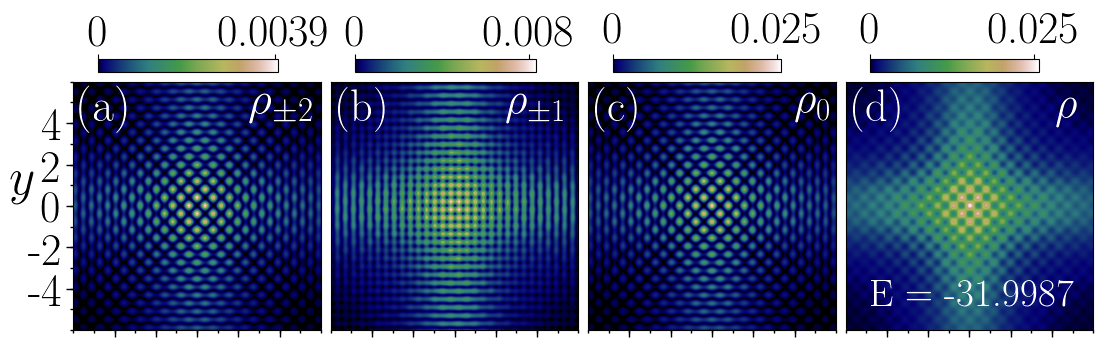}
\includegraphics[trim = 0cm 0cm 0cm 0cm, clip,width=1\linewidth,clip]
                {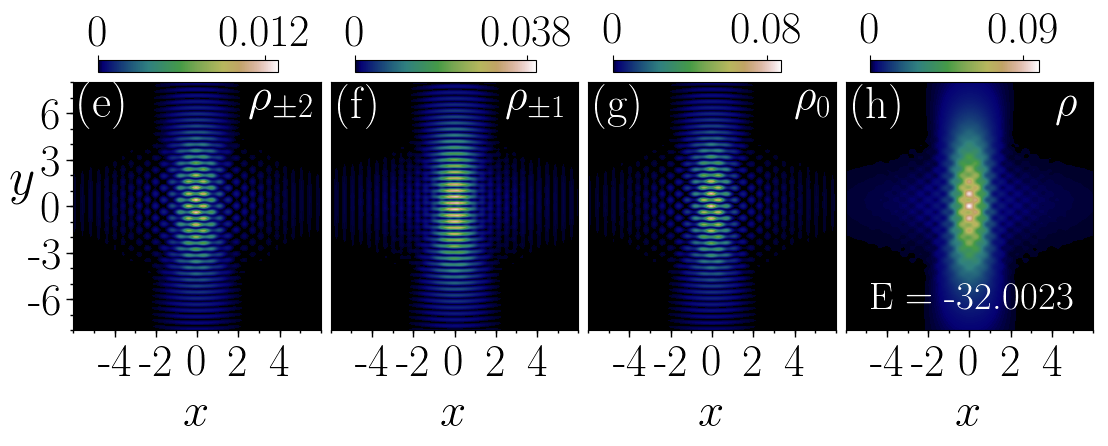}
\caption{(Color online) Contour plot of density of a square-lattice
soliton of  components (a) $j = \pm 2$, (b) $j = \pm 1$, (c) $j = 0$, and (d) total density 
with $c_0 = -0.5$, $c_1 = 0.025$, $c_2 = -0.25$, (polar phase) $\gamma = 4$, $E=-31.9987$; 
the same of a superstripe soliton for the same parameters in (e)-(h) with $E=-32.0023$.} 
\label{fig6}
\end{center}
\end{figure}

In the polar phase, with $c_0 = -0.5$, $c_1 = 0.025$, 
$c_2 = -0.25$, and $\gamma\gtrapprox4$, we get the  same four quasi-degenerate solitons as in 
the cyclic phase discussed above. Two of these, square-lattice soliton with energy $-31.9987$ 
and superstripe soliton with energy $-32.0023$ are shown in 
Figs.~\ref{fig6}(a)-(d) and (e)-(h), respectively.  
The superstripe soliton has a square-lattice type spatial modulation superposed  on stripes 
in components $j=\pm 2$ and 0 and a stripe modulation in component $j=\pm 1$ whereas total
density has a square-lattice type pattern. The square-lattice soliton is quite 
similar to same of Figs. \ref{fig4}(a)-(d) and \ref{fig5}(e)-(h). However,
the superstripe soliton  of Figs.~\ref{fig6}(e)-(h) has now acquired a square-lattice pattern 
in total density quite similar to a superstripe soliton of an SO-coupled spin-1 spinor BEC
for $\gamma=8$  \cite{adhikari2021multiring} in both ferromagnetic and polar phases. 
The stripe soliton with energy $-32.0044$ and  multi-ring soliton with energy $-32.0009$ are not shown here.

\subsection{Dynamical Stability}
\label{dyn_inst}

\begin{center}
\begin{figure}[t]
\includegraphics[trim = 0cm 0cm 0cm 0cm, clip,width=1\linewidth,clip]
                {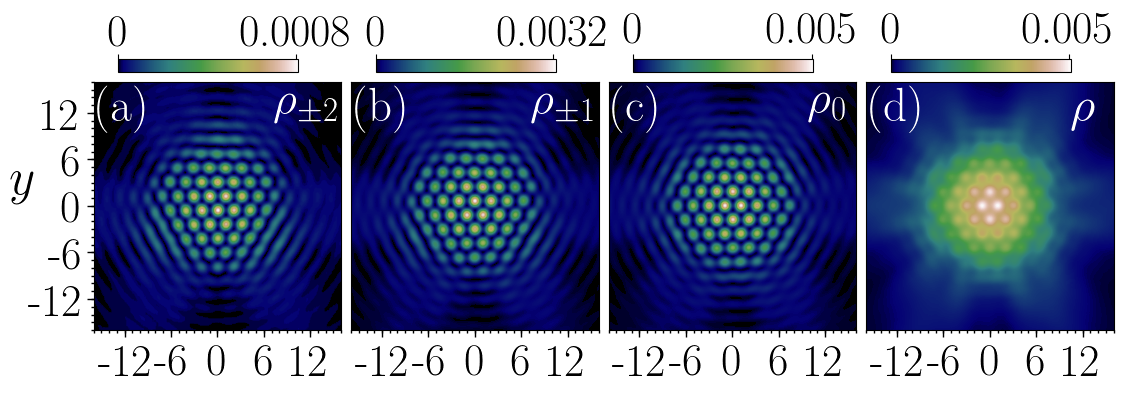}
\includegraphics[trim = 0cm 0cm 0cm 0cm, clip,width=1\linewidth,clip]
                {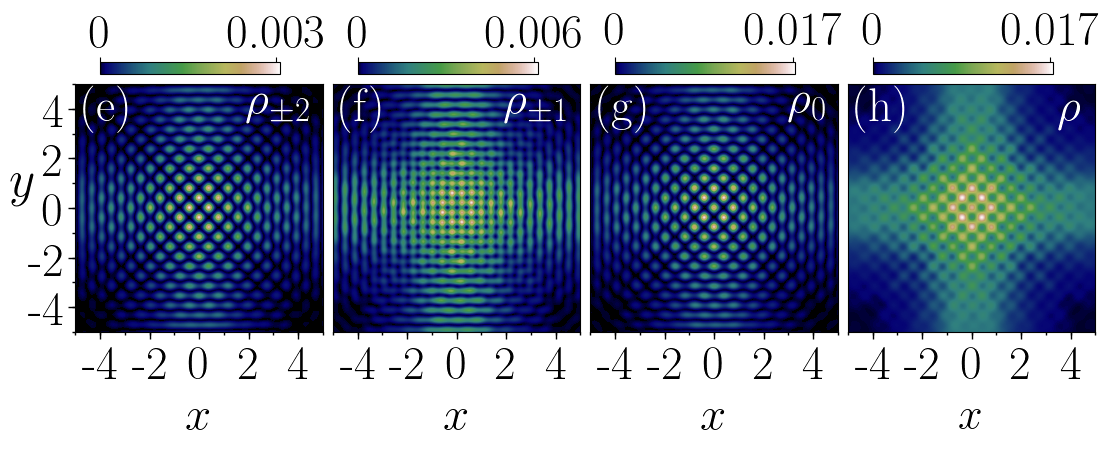}
\caption{(Color online) Contour plot of component densities of the  triangular-lattice soliton 
of Figs. 7(a)-(d) for 
components (a) $j = \pm 2$, (b) $j = \pm 1$, (c) $j = 0$, and (d) total density after $100$ units of time; 
the same of the  square-lattice soliton of Figs. 8(a)-(d) for components (e) $j = \pm 2$, (f) $j = \pm 1$, (g) $j = 0$, and (h) 
total density, {after real-time simulation over 100 units of time}. The initial state used in real-time propagation is obtained by adding a random noise 
(\ref{rand_noise}) to the stationary-state imaginary-time solutions shown in Figs. \ref{fig3}(a)-(d) for (a)-(d) and 
Figs. \ref{fig4}(a)-(d) for (e)-(h).
}
\label{figure11}
\end{figure}
\end{center}

We confirm the dynamical stability of the stationary states of the SO-coupled spin-2 BEC
discussed in Sec. \ref{section3a}-\ref{section3c} via a real-time propagation over an extended period
of time up to $t = 500$. In addition to this, we have also tested the stability of these solutions by 
adding an initial random noise $\delta {\phi_{j}}^{\rm noise}$ to the respective order parameters
at $t = 0$ and then studying their real-time dynamics. We consider the random noise as 
\begin{equation}
\delta {\phi_{j}}^{\rm noise}(x,y) = {10}^{-3}  \sqrt{{\cal N}_j} {\rm R}_g(x,y) e^{i{\rm R}_u(x,y)},
\label{rand_noise}
\end{equation}
where ${\cal N}_j = \int \rho_j(x,y)d{\bf r}$.
The amplitude of this noise is randomized by random numbers $R_g(x,y)$ which follow the Gaussian 
distribution, whereas phase of the noise is randomized by $R_u(x,y)$ which follows a 
uniform probability distribution over the interval $[0,2\pi]$. 
As an illustration, we consider the triangular-lattice soliton of Figs.~\ref{fig3}(a)-(d) and the 
square-lattice soliton of Figs.~\ref{fig4}(a)-(d). At $t= 0$, $\delta {\phi_{j}}^{\rm noise}$ is added
to the respective order-parameters and the resultant order parameters are considered initial solutions to
Eqs.~(\ref{cgpet3d-1})-(\ref{cgpet3d-3}), which are now solved (evolved) in real-time up to $t=100$. 
The resultant component and total densities at $t = 100$ are displayed in 
Figs.~\ref{figure11}(a)-(d) 
and \ref{figure11}(e)-(h), respectively. The periodic density patterns survive with slightly different 
peak densities compared to the $t = 0$ solutions, demonstrating the dynamical stability of the 
solitons.
\subsection{Bifurcations}
\label{bifur}
\begin{figure}[H]
\begin{center}
\includegraphics[width=7.3cm ]    
                {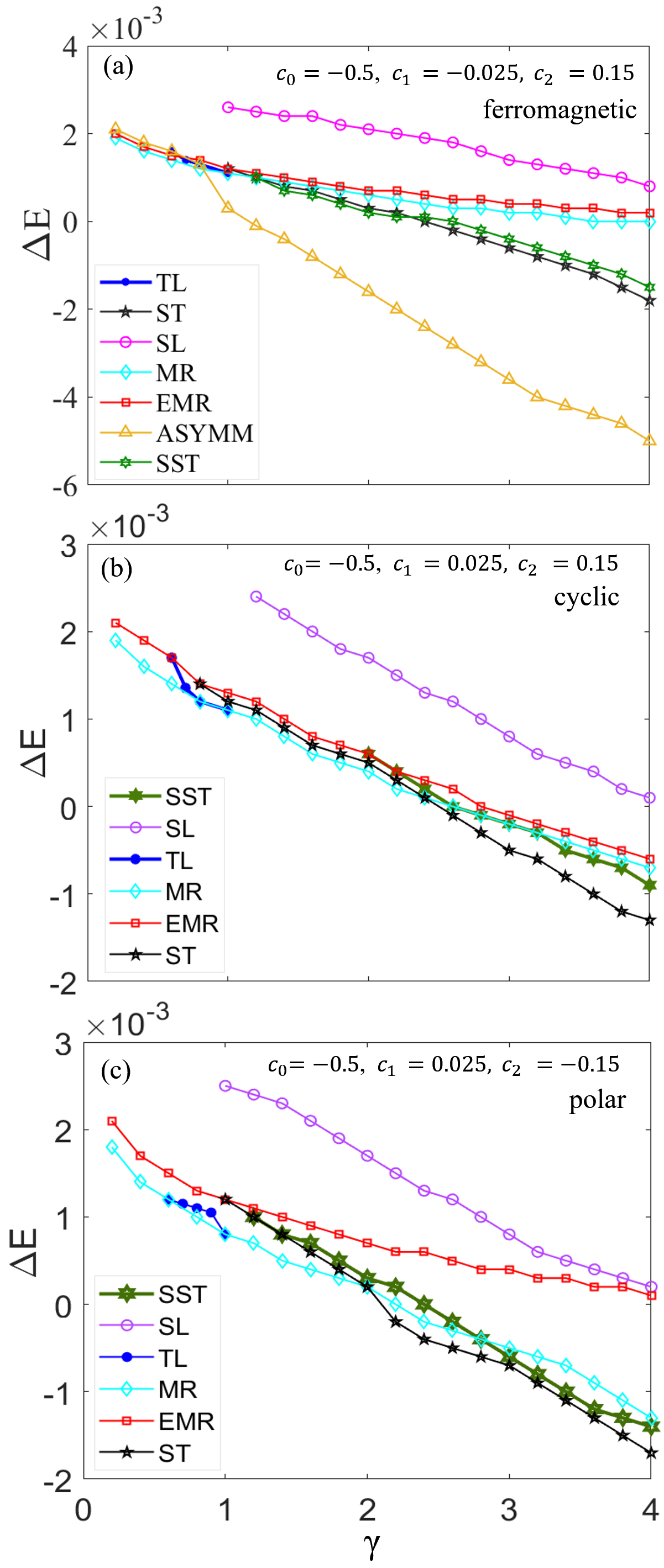}
\caption{(Color online) (a) The bifurcation diagram in the ${\Delta}E$-$\gamma$ plane for the
ferromagnetic phase corresponding to a cut in the phase-diagram \ref{fig1x}(a) at $c_2 =  0.15$. The same for 
the cyclic phase is shown in (b) by taking a cut at $c_2 = 0.15$ in the phase-diagram \ref{fig1x}(b). In the 
polar domain of phase diagram \ref{fig1x}(b), a cut is taken at $c_2 = -0.15$, and bifurcation picture is shown
in (c). The $\Delta E = E+2\gamma^2$, where $E$ is the energy of the state and $-2\gamma^2$ is the single particle solution's energy, corresponding to various quasi-degenerate states are plotted by using different 
symbols as well as different colours. SST corresponds to the super-stripe similar to the state shown
in \ref{fig6}(e)-(h), SL corresponds to the square-lattice similar to the state shown
in \ref{fig4}(a)-(d), TL corresponds to the triangular-lattice similar to the state shown
in \ref{fig3}(a)-(d), MR corresponds to the (-2, -1, 0, +1, +2)-type multi-ring solution similar to the state 
shown in \ref{fig1}(a)-(d), EMR corresponds to the  (-1, 0, +1, +2, +3)-type excited-state multi-ring solution, 
ST corresponds to the stripe solution similar to the state shown in \ref{fig5}(a)-(d), ASYMM corresponds to 
the circularly-asymmetric solution similar to the state shown in \ref{fig2}(f)-(j).} 
\label{bifurfaction_fig}
\end{center}
\end{figure}

In the non-interacting system, various solutions are completely degenerate, whereas on the introduction 
of interactions, solutions of Eqs.~(\ref{cgpet3d-1})-(\ref{cgpet3d-3}) exhibit a bifurcating 
behaviour. As the energies of these solutions are very close, to make the nature of these bifurcations clear, 
we calculate the difference $\Delta E$ between total energy of the solution 
and the single-particle solution's energy, i.e., $-2\gamma^2$ as discussed in Sec. \ref{section2b}. 
A cut is now considered in the phase-diagram \ref{fig1x}(a) at an appropriate $c_2$, say  $c_2 = 0.15$, 
for the ferromagnetic phase, and $\Delta E$ as a function of SO coupling strength $\gamma$ is evaluated for the
various solutions. Similarly, a cut at $c_2 = 0.15$ for the cyclic phase in \ref{fig1x}(b) and at $c_2 = -0.15$ for the polar 
phase in \ref{fig1x}(b) are considered. The resultant bifurcation plots showing $\Delta E$ as a
function of the SO-coupling strength $\gamma$ are shown in Figs. \ref{bifurfaction_fig}(a)-(c) for the
three magnetic phases. Bifurcation points agree with the critical points in the phase diagrams shown in 
Figs. \ref{fig1x}(a)-(b).

\subsection{Moving  $(-2,-1,0,+1,+2)$-type soliton}\label{section3e}

\begin{figure}[t]
\begin{center}
\includegraphics[trim = 0cm 0cm 0cm 0cm, clip,width=\linewidth, clip]
                {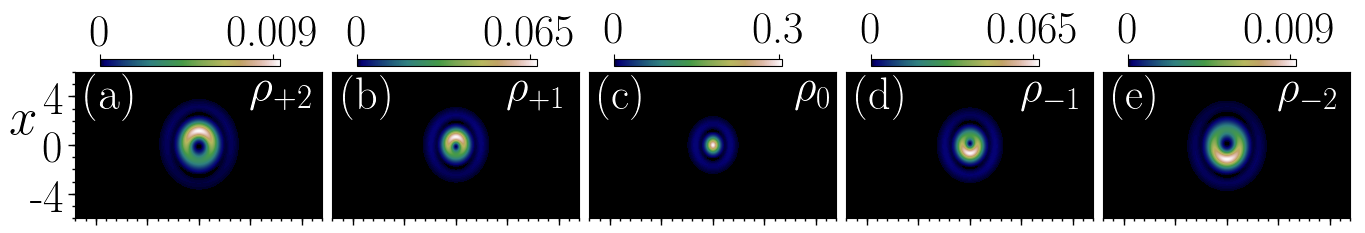}
\includegraphics[trim = 0cm 0cm 0cm 0cm, clip,width=\linewidth,clip]
                {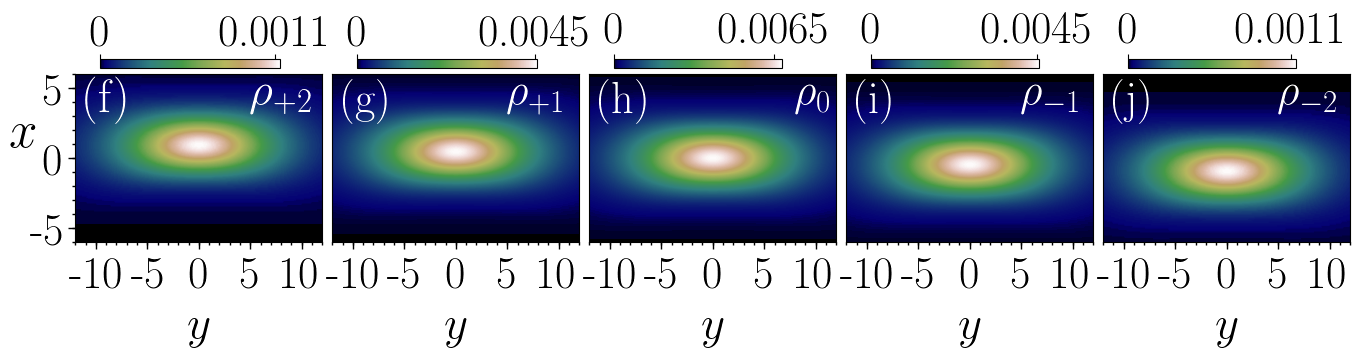}
\caption{(Color online) Contour plot of density of  components 
(a) $j = +2$, (b) $j = +1$, (c) $j = 0$, (d) $j = -1$, and (e) $j = -2$ 
with $c_0 = -2.5$, $c_1 = -0.025$, $c_2 = 0.25$, and 
$\gamma = 0.5$ moving with $v = 0.03$ along $+y$;  the
same densities for velocity $v=0.1$ in (f)-(j).}
\label{fig7}
\end{center}
\end{figure}

The SO coupling breaks the Galilean invariance of the mean-field model 
of the spinor BECs \cite{Gautam_spin2}. Explicitly, considering the Galilean 
transformation $x'=x, y'=y-v t, t'=t$, where $v$ is the relative velocity 
along $y$-axis of primed coordinate system with respect to unprimed coordinate 
system, along with the following transformation of the wave function $\psi^\prime \to  \phi$
\begin{equation}
\phi_j(x,y,t) = {\psi_j}'(x',y',t')e^{iv y'+iv^2 t'/2},
\label{movingansatz}
\end{equation}
we get from Eqs. (\ref{cgpet3d-1})-(\ref{cgpet3d-3})
\begin{subequations}
\begin{align}
i \partial_{t^\prime}  \psi^{\prime}_{\pm 2} =
 & \mathcal{H} \psi^{\prime}_{\pm 2}
 + c_0 {\rho}  \psi^{\prime}_{\pm 2}+ c_1 (F^{\prime}_{\mp} 
 \psi^{\prime}_{\pm 1} \pm 2 F^{\prime}_{z} \psi^{\prime}_{\pm 2}) 
\nonumber\\ +& \textstyle\frac{c_2}{\sqrt 5}
\Theta \psi^{\prime}_{\mp  2}
  -i\gamma \partial_\mp
\psi^{\prime}_{\pm 1} 
+{\gamma \psi^{\prime}_{\pm 1}v},\label{moving1}\\
i \partial_{t^\prime}  \psi^{\prime}_{\pm 1} 
=& \mathcal{H} \psi^{\prime}_{\pm 1} + c_0 {\rho}\psi^{\prime}_{\pm 1} 
+ c_1 \left(\textstyle\sqrt{\frac{3}{2}} F^{\prime}_{\mp} \psi^{\prime}_{0} 
+F^{\prime}_{\pm} \psi^{\prime}_{\pm 2} \right. \nonumber\\
 \pm&\left.
F^{\prime}_{z}\psi^{\prime}_{\pm 1}\right)
-\textstyle \frac{c_2}{\sqrt{5}}\Theta \psi^{\prime}_{\mp 1}
- i \gamma\textstyle \sqrt{\frac{3}{2}}
\partial _\mp
 \psi^{\prime}_{ 0}
\nonumber\\ 
-&i\gamma \partial_\pm   \psi^{\prime}_{\pm 2} 
+{\gamma\left(\textstyle\sqrt{\frac{3}{2}}\psi^{\prime}_0+\psi^{\prime}_{\pm 2}
\right)v},
\label{moving2}\\
i{\partial_{t^\prime} \psi^{\prime}_0}
 =& \mathcal{H} \psi^{\prime}_0 + c_0 {\rho}\psi^{\prime}_0 
+ c_1 {\textstyle \sqrt{\frac{3}{2}}}(F_{-} \psi^{\prime}_{-1} 
+  F_{+} \psi^{\prime}_{+1})  
  \nonumber\\+
& \textstyle  \frac{c_2}{\sqrt{5}}\Theta
\psi^{\prime}_{0}-i \textstyle \sqrt{\frac{3}{2}}\gamma
 \partial_+(\psi^{\prime}_{+1}+\psi^{\prime}_{-1} )  
 \nonumber\\
+&    {\gamma\textstyle \sqrt{\frac{3}{2}}\left(\psi^{\prime}_{+1}
+\psi^{\prime}_{-1}\right)v}\label{moving3},
\end{align}
\end{subequations}
where $\partial_{\pm}= (\partial_{x^\prime } 
\pm
i\partial_{y^\prime })$.
These equations are distinct from Eqs.~(\ref{cgpet3d-1})-(\ref{cgpet3d-3}) 
indicating a breakdown of the Galilean invariance. For an SO-coupled spin-2 
BEC, the moving solitons are the stationary solutions of 
Eqs.~(\ref{moving1})-(\ref{moving3}) multiplied by a factor of $e^{ivy}$. 
The structure of the moving soliton depends on the magnitude 
as well as direction of velocity.
Here we study the fate of a moving  $(-2,-1,0,+1,+2)$-type multi-ring soliton 
by solving Eqs.~(\ref{moving1})-(\ref{moving3}) numerically, for small 
SO-coupling strength, as the velocity is increased. For example, considering 
$c_0 = -2.5, c_1 = -0.025, c_2 =$  0.25, $\gamma = 0.5$ with 
(a) $v = 0.03$ and (b) $v = 0.1$, the component densities of the moving 
solitons are shown in Figs.~\ref{fig7}(a)-(e) and (f)-(j), 
respectively. The structure of the moving soliton at two different velocities are 
quite distinct as can be seen in Fig. \ref{fig7}.
With the  increase of velocity along $y$ axis, component 
phase-singularities move along $x$ axis to region of low density away from the center, 
resulting in component densities without any vortex core at velocity  $v=0.1$. 
For $c_0 = -2.5, c_1 = -0.025, c_2 = 0.25$, $\gamma = 0.5$, a self-trapped 
moving soliton with $v > 0.3$ does not exist.

\begin{figure}[t]
\includegraphics[trim = 0cm 0cm 0cm 0cm, clip,width=\linewidth, height=4 cm,
clip]
{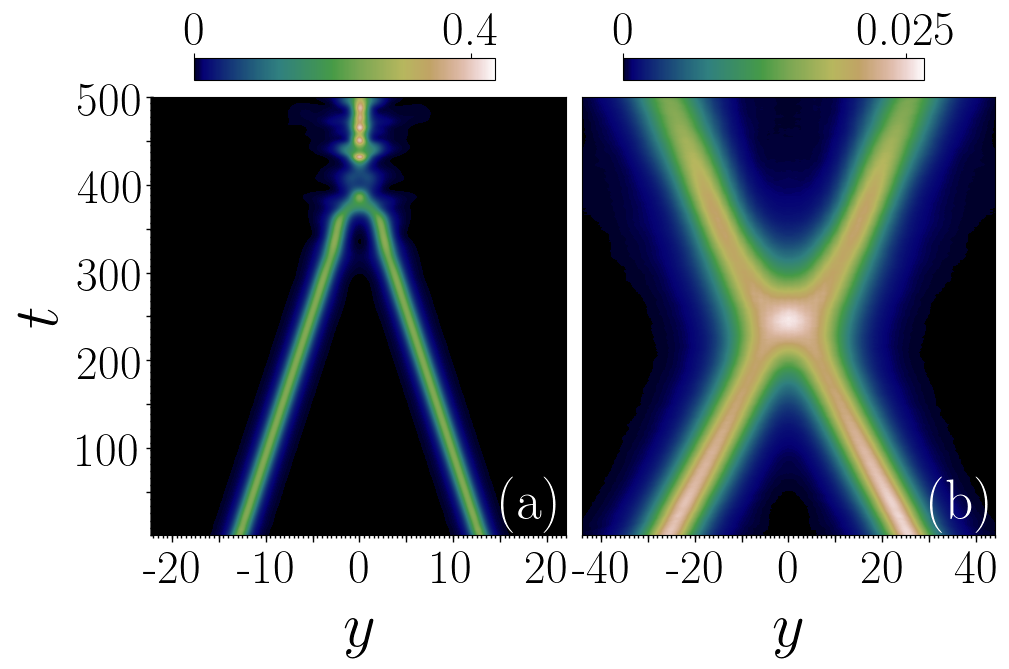}
\caption{(Color online)
Contour plot of total density $\rho(0,y,t)$ as a function 
of $y$ and $t$ during the head-on collision between the two solitons
with $c_0 = -2.5$, $c_1 = -0.025$, $c_2 = 0.25$, $\gamma =0.5$ moving with velocity  
(a) $|v| =\pm  0.03$ and (b) $|v|=\pm 0.1$ along $y$ axis in opposite direction.}
\label{fig9}
\end{figure}

%The dynamics of the bright-soliton moving with $v = \pm 0.03$ for the 
%condensate with $c_0 = -2.5, c_1 = -0.025, c_2 = -0.25$, $\gamma = 0.5$ 
%is shown in Fig.~\ref{fig8}(a) and (b).

%\begin{figure}[t]
%\begin{center}
%\includegraphics[trim = 0cm 0cm 0cm 0cm, clip,width=\linewidth,clip]{fig8.png}
%\caption{(Color online) The contour plot of $\rho(0,y,t)$ as a function of
%$y$ and $t$ corresponding to the soliton with $c_0 = -2.5$, 
%$c_1 = -0.025$, $c_2 = 0.25$, $\gamma =0.5$ and moving with 
%(a) $v = -0.03$ and (b) $v = 0.03$.}
%\label{fig8}
%\end{center}
%\end{figure}

We have also studied the head-on collision of these solitons. At low 
velocities, the collision is inelastic while the solitons come close 
to each other interact and form a bound entity and never come out.
On the other hand,  at large initial velocities  the collision is quasi elastic; 
in this case the solitons tend to pass through 
each other without any change of velocity. For example, the head-on collision 
between the solitons
moving with $|v| = 0.03$ and $c_0 = -2.5, c_1 = -0.025, c_2 = 0.25$, 
$\gamma = 0.5$, is shown in Fig.~\ref{fig9}(a) through a contour plot of
time evolution of total density 
$\rho(0,y,t)$ in the $t-y$ plane. 
Similarly, a   quasi-elastic collision  between two solitons moving 
with velocity $|v| = 0.1$ is shown in Fig.~\ref{fig9}(b). 
The collision dynamics is 
consistent with the similar observations for two SO-coupled spin-1 
BECs \cite{gautam2017vortex}.

%%%%%%%%%%%%%%%%%%%%%%%%%%%%%%%%%%%%%%%%%%%%%%%%%%%%%%%%%%%%%%%%%%%%%%%%%%%%%%%%
%%%%%%%%%%%%%%%%          Summary and Conclusions    
%%%%%%%%%%%%%%%%%%%%%%%%%%%%%%%%%%%%%%%%%%%%%%%%%%%%%%%%%%%%%%%%%%%%%%%%%%%%%%%%

\section{Summary and Conclusion}

We have demonstrated  the emergence of various self-trapped stable 
solitons with supersolid-like crystallization in a quasi-2D  
SO-coupled spin-2 BEC employing  analytic consideration and numerical solution
of the underlying   mean-field GP equation.  The minimization of interaction and 
SO-coupling energies leads to the permissible winding-number combinations
for axisymmetric solitons. In the absence of interactions, we consider the 
eigenfunctions of the single-particle Hamiltonian to construct the order 
parameters consistent with multi-ring, stripe, triangular-, and square-lattice
density profiles. In the presence of (attractive) interactions, we find  
that various  types of solitons with spatially-periodic modulation in density appears,
including the ones inferred from a study of the eigenfunctions of the single-particle
Hamiltonian,  due to an interplay of SO-coupling and interactions. 

The ground state for  a small SO-coupling strength  ($\gamma \approx 0.5$) is a 
radially symmetric multi-ring soliton for weakly-ferromagnetic, cyclic and 
polar interactions, whereas for a sufficiently strong-ferromagnetic interaction, 
circularly-asymmetric soliton emerges as the ground state. For intermediate SO-coupling 
strengths ($\gamma \approx 1$), in addition to the  axisymmetric soliton, there could 
exist a triangular-lattice soliton with a hexagonal crystallization of 
matter in the soliton, explicit in both component and total densities. On increasing the SO-coupling 
further, one could have five quasi-degenerate solitons, e.g. a  multi-ring soliton,  
a square-lattice soliton,  a stripe soliton and a superstripe soliton,
in all the magnetic phases, and also circularly-asymmetric soliton in the ferromagnetic phase.
The quasi-degeneracy between the states is in general lifted with either a decrease in  the 
SO-coupling strength $\gamma$ or an increase in the attractive spin-independent interaction strength $|c_0|$.
We  also introduced the Galilean-transformed 
model to study the moving solitons and the  head-on collision dynamics
between two such solitons. A head-on collision between the two solitons is inelastic 
at low velocities and   the two solitons  can form  a bound entity. At large 
velocities the collision is quasi elastic, and the  solitons pass through each other
without a substantial change of velocity.

\begin{acknowledgments}

SG acknowledges the support of the Science \& Engineering Research Board (SERB),
Department of Science and Technology, Government of India under the Project
ECR/2017/001436. SKA acknowledges partial support
by the CNPq (Brazil) grant 301324/2019-0, and by the
ICTP-SAIFR-FAPESP (Brazil) grant 2016/01343-7.

\end{acknowledgments}

%%%%%%%%%%%%%%%%%%%%%%%%%%%%%%%%%%%%%%%%%%%%%%%%%%%%%%%%%%%%%%%%%%%%%%%%%%%%%%%%
%%%%%%%%%%%%%%%%            Appendix
%%%%%%%%%%%%%%%%%%%%%%%%%%%%%%%%%%%%%%%%%%%%%%%%%%%%%%%%%%%%%%%%%%%%%%%%%%%%%%%%
\begin{widetext}
\section*{Appendix}
For a spin-2 BEC, the spin-dependent interaction energy is given as \cite{spinor_review}
\begin{equation}
\label{spindependent}
E_{\rm int}=  \int\Big(\frac{c_1}{2}|{\bf F}|^2 + 
\frac{c_2}{2}|\Theta|^2 \Big) rdr d\theta 
\end{equation}

%The phase dependent terms from $|{\bf F}|^2$ are
%\begin{eqnarray}
%E_{int}^1&=&c_1\int \Re(2\sqrt6 \phi_0 {\phi_1}^{*2}\phi_2 + 6{\phi_0}^{*2}
%\phi_1\phi_{-1}+2\sqrt6{\phi_{0}^*} {\phi_{1}^*}\phi_2\phi_{-1}+
%2\sqrt6{\phi_{0}^*} {\phi_{1}}\phi_{-1}^*\phi_{-2} +4\phi_{1}^
%*\phi_2\phi_{-1}^*\phi_{-2}+
%2\sqrt{6}\phi_0\phi_{-1}^{*2}\phi_{-2})dr d\theta
%\end{eqnarray}
Using  {\em ansatz} (\ref{anstaz}), the contribution of 
phase dependent terms in the spin dependent interaction energy 
(\ref{spindependent}) can be written as
\begin{eqnarray}
\label{INTERENERGY}
E_{\rm int}^{\rm phase} &=& 2\sqrt{6}c_1\textstyle \int R_0 R_{+1}^2 R_{+2} r dr\int
\cos[(w_0-2w_{+1}+w_{+2})\theta+(\alpha_0-2\alpha_{+1}+\alpha_{+2})]d\theta+
2\Big(3c_1-\frac{c_2}{5}\Big)\int R_0^{2} R_{+1} R_{-1}r dr \nonumber\\ &\times& 
\textstyle  \int \cos[(w_{+1}+w_{-1}-2w_0)\theta +(\alpha_{+1}+\alpha_{-1}-2\alpha_0)]
d\theta+2\sqrt{6}c_1\int R_0 R_{-1}^2 R_{-2}r dr
\int \cos[(w_0+w_{-2}-2w_{-1})\theta \nonumber\\&+&
 \textstyle 
(\alpha_0+\alpha_{-2}-2\alpha_{-1})]d\theta
\nonumber\\&+&  
2\sqrt{6}c_1\int R_0 R_{+1} R_{+2} R_{-1}r dr 
\int \cos[(w_{+2}+w_{-1}-w_0-w_{+1})\theta+ (\alpha_{+2}+\alpha_{-1}-\alpha_0-\alpha_{+1})]
d\theta \nonumber\\&+& \textstyle 2c_1\sqrt{6}\int R_0 R_{+1} R_{-1} R_{-2} r dr
 \int \cos[(w_{+1}+w_{-2}-w_0-w_{-1})\theta+
(\alpha_{+1}+\alpha_{-2}-\alpha_0-\alpha_{-1})]d\theta \nonumber\\&+&
\textstyle 4(c_1-\frac{c_2}{5})\int R_{+1} R_{+2} R_{-1} R_{-2}r dr
\int \cos[(w_{+2}+w_{-2}-w_{+1}-w_{-1})+(\alpha_{+2}+\alpha_{-2}-\alpha_{1}-\alpha_{-1})]
d\theta\nonumber\\&+&
\textstyle 
\frac{2c_2}{5}\int R_{+2} R_{-2} R_0^{2}r dr \int 
\cos[(2w_0-w_{+2}-w_{-2})\theta+(2\alpha_0-\alpha_{+2}-\alpha_{-2})]d\theta\, .
\end{eqnarray}
A typical $\theta$-dependent term in Eq.~(\ref{INTERENERGY}) can be written as {\cite{han-double}} 
\begin{equation} 
\int_{0}^{2\pi} \cos(w_s\theta + \alpha_s)d\theta = 
\frac{\sin(2\pi w_s + \alpha_s)}{w_s} -\frac{\sin{\alpha_s}}{w_s}, 
\label{cos_int}
\end{equation}
where $w_s$ and $\alpha_s$, represent any of the linear combinations 
of $w_j$'s and $\alpha_j$'s appearing as arguments of cosine, respectively. 
As $w_s$ can only be an integer including zero, the absolute 
value of integral (\ref{cos_int}) is $2\pi$ if $w_s = 0$ and $\alpha_s$ 
is an integer multiple of $\pi$. The exact values of $\alpha_s$ has 
to be determined by minimizing energy (\ref{INTERENERGY}) with 
$w_s = 0$. The permitted independent winding number relations thus are
\begin{eqnarray}
\label{winding-interaction}
w_{+1} + w_{-1} -2 w_0 = 0,\quad
w_{+2} + w_{-2} - w_{+1} - w_{-1} = 0,\quad
w_0 + w_{-2} - 2w_{-1} = 0.
\end{eqnarray}

The energy contribution from SO-coupling terms, obtained 
by using { ansatz} (\ref{anstaz}), is
\begin{align}\label{soc_energy}
E_{\rm so} &=\textstyle \int dx dy \sum_j  \phi_j^*\Gamma_j \nonumber \\
&= \gamma \int {d\bf{r}} \left[R_{+2} e^{i[(w_{+1}-w_{+2}-1)
\theta+(\alpha_{+1}-\alpha_{+2})]}\Big[\frac{\partial R_{+1}}{\partial r} 
+ w_{+1} \frac{R_{+1}}{r} \Big]+R_{-1}
e^{i[(w_{-2}-w_{+1}-1)\theta+(\alpha_{-2}-\alpha_{-1})]}
\Big[\frac{\partial R_{-2}}{\partial r}+\frac{w_{-2} R_{-2}}{r}\Big] \right.
\nonumber\\
& \left. +R_{+1}e^{i[(w_{+2}-w_{+1}+1)
\theta+(\alpha_{+2}-\alpha_{+1})]}\Big[\frac{-\partial R_{+2}}{\partial r}+
\frac{w_{+2} R_{+2}}{r}\Big]+R_{-2} 
e^{i[(w_{-1}-w_{-2}+1)\theta+(\alpha_{-1}-\alpha_{-2})]}
\Big[\frac{-\partial R_{-1}}{\partial r} + w_{-1} \frac{R_{-1}}{r} \Big] \right]
\nonumber\\
&+\sqrt{\frac{3}{2}}\gamma \int d {\bf r}  \left[ R_{+1} 
e^{i[(w_0-w_{+1}-1)\theta+(\alpha_0-\alpha_{+1})]}
\Big[\frac{\partial R_0}{\partial r}+\frac{w_0 R_0}{r}\Big]
+R_{-1} e^{i[(w_0-w_{-1}+1)\theta+(\alpha_0-\alpha_{-1})]}
\Big[\frac{-\partial R_0}{\partial r}+\frac{w_0 R_0}{r}\Big]  \right.\nonumber\\&
\left.+R_0 e^{i[(w_{-1}-w_0-1)\theta+(\alpha_{-1}-\alpha_0)]}
\Big[\frac{\partial R_{-1}}{\partial r}+\frac{w_{-1} R_{-1}}{r}\Big]
+R_0 e^{i[(w_{+1}-w_0+1)\theta+(\alpha_{+1}-\alpha_0)]}
\Big[\frac{-\partial R_{+1}}{\partial r}+\frac{w_{+1} R_{+1}}{r}\Big]
\right],
\end{align}
\end{widetext}
where ${\bf r}\equiv \{x,y\}\equiv \{r,\theta\}$.
Again, a minimization of $E_{\rm so}$ requires that 
\begin{eqnarray}
\label{windingsoc}
 w_{+2}-w_{+1}+1 = 0,
\quad w_{+1}-w_0+1=0, \\
\quad w_{-2}-w_{-1}-1=0,
\quad w_{-1}-w_0-1=0,\label{windingsoc2}
\end{eqnarray}
and linear combinations of $\alpha_j$'s appearing in Eq.~(\ref{soc_energy})
are integer multiple of $\pi$. The winding number relations in 
Eq.~(\ref{winding-interaction}) are not independent as all can be obtained
from winding number relations in Eqs.~(\ref{windingsoc}) and (\ref{windingsoc2}).

%%%%%%%%%%%%%%%%%%%%%%%%%%%%%%%%%%%%%%%%%%%%%%%%%%%%%%%%%%%%%%%%%%%%%%%%%%%%%%%
%%%%%%%%%%%%             Acknowledgements                          %%%%%%%%%%%%
%%%%%%%%%%%%%%%%%%%%%%%%%%%%%%%%%%%%%%%%%%%%%%%%%%%%%%%%%%%%%%%%%%%%%%%%%%%%%%%

\end{document}